\documentclass[letterpaper11pt]{article}
\usepackage[utf8]{inputenc}
\usepackage[T1]{fontenc}
\usepackage{hyperref}
\usepackage[margin=1in]{geometry}
\usepackage{amssymb}
\usepackage{amsmath}
\usepackage{dsfont}
\usepackage{mathtools}
\usepackage{authblk}
\usepackage{xcolor}
\usepackage{dsfont}
\usepackage{subcaption}
\usepackage{wrapfig}

\usepackage{natbib}
\usepackage{multido} 
\usepackage{soul}

\title{\LARGE \textbf{Variance-reduced extreme value index estimators\\ using control variates in a semi-supervised setting}}
\author[1,2]{Louison Bocquet-Nouaille}
\author[1,2]{Jérôme Morio}
\author[2]{Benjamin Bobbia}
\affil[1]{ONERA/DTIS, Université de Toulouse, F-31055 Toulouse}
\affil[2]{Fédération ENAC ISAE-SUPAERO ONERA, Université de Toulouse, 31000 Toulouse}
\date{}

\newtheorem{definition}{Definition}
\newtheorem{theorem}{Theorem}
\newtheorem{proposition}{Proposition}
\newtheorem{corollary}{Corollary}[theorem]

\newcommand{\E}{\mathbb{E}}
\newcommand{\1}{\mathds{1}}
\newcommand{\var}{\text{Var}}
\newcommand{\cov}{\text{Cov}}
\newcommand{\corr}{\text{Corr}}

\begin{document}

\maketitle

\noindent\rule{\linewidth}{0.4pt}
The estimation of the Extreme Value Index (EVI) is fundamental in extreme value analysis but suffers from high variance due to reliance on only a few extreme observations. We propose a control variates based transfer learning approach in a semi-supervised framework, where a small set of coupled target and source observations is combined with abundant unpaired source data. By expressing the Hill estimator of the target EVI as a ratio of means, we apply approximate control variates to both numerator and denominator, with jointly optimized coefficients that guarantee variance reduction without introducing bias. We show theoretically and through simulations that the asymptotic relative variance reduction of the transferred Hill estimator is proportional to the tail dependence between the target and source variables and independent of their EVI values. Thus, substantial variance reduction can be achieved even without similarity in tail heaviness of the target and source distributions. The proposed approach can be extended to other EVI estimators expressed with ratio of means, as demonstrated on the moment estimator. The practical value of the proposed method is illustrated on multi-fidelity water surge and ice accretion datasets.
\noindent\rule{\linewidth}{0.4pt}

\section{Introduction}


A key parameter in Extreme Value Theory (EVT) is the Extreme Value Index (EVI), which characterizes the behavior of the distribution's tail, notably the frequency and magnitude of extreme events. Estimating the EVI is essential to compute other quantities, such as extreme quantiles.

Classical EVT approaches assume a large dataset of size $n$, but only use the top $k<<n$ observations representing extremes to study the distribution’s tail. Choosing the number of extremes $k$ involves a crucial bias–variance trade-off: taking $k$ too large introduces bias by including non-extreme observations, while taking $k$ too small results in estimates with high variance due to insufficient data. In addition, datasets are often limited in practice, making both the sample size $n$ and the number of extremes $k$ small. As a result, each observation influences the estimates more strongly, which increases their variance and leads to less reliable results. \\

Many variance reduction methods exist in different frameworks \citep{asmussen_variance-reduction_2007}. Yet, in extreme value analysis, most studies focused on bias correction \citep{caeiro_minimum_2020, gomes_new_2016, cai_bias_2013}. Given the bias–variance trade-off, variance reduction constitutes a natural and complementary strategy to bias correction for improving the accuracy of EVI estimates.

In this work, a transfer learning approach for variance reduction \citep{zhu_recent_2025} is considered. The idea of transfer learning is to leverage information from another dataset, called the source, to enhance estimations on the target dataset. This approach is similar to how humans learn: someone who knows how to play the piano will find it easier to learn the violin than someone with no musical experience. In practice, the dataset considered is composed of a small set of coupled target and source samples $(Y^T_i, Y^S_i)_{i=1\dots n}$, and a larger set of additional unpaired source samples $(Y^S_i)_{i=n+1\dots n+m}$. This setting assumes that new samples cannot be acquired. The transfer success relies on the dependence between target and source, combined with the abundance of source data. This framework is often referred to as the "semi-supervised" framework, for its similarity with machine learning problems where only a few observations are labeled in larger datasets. Although only one source is considered in this work, the framework can be extended to multiple sources and is closely related to data fusion, where multiple datasets are integrated to enhance inference.

The semi-supervised setting appears in a wide range of applications. The target may correspond to real-world data, while the source comes from a simulation model. Alternatively, the target could come from a more accurate, high-fidelity model, while the source comes from a cheaper, lower-fidelity model. This is known as multi-fidelity modeling, and it is particularly suited for transfer learning \citep{peherstorfer_survey_2018}. Another relevant setting involves source and target data of similar fidelity, but collected at different locations. For example, wind measurements from two nearby weather stations, where the source station may offer higher sampling resolution or a longer observation history. In each case, the richer source data helps improve estimation on the target. \\

A few recent studies have employed transfer learning to reduce the variance of extreme value estimations in a semi-supervised setting. For instance, \citep{kim_parametric_2024} apply a parametric Multi-Fidelity Monte Carlo (MFMC) method to estimate parameters of the Gumbel distribution and exceedance probabilities. This method, also known as approximate control variates (ACV) \citep{gorodetsky_generalized_2020}, is the one adopted in the present work and further detailed in Section \ref{sec:CV}. Similarly, \citep{ahmed_extreme_2025} introduce variance-reduced estimators for the EVI, based on the maximum likelihood estimator (MLE), and for extreme quantiles. The authors propose the following approach: the source variable is transformed to have a known EVI $g$, which is estimated with the MLE as $\widehat{g}$, to refine the target EVI estimation with the difference $\widehat{g}-g$. In earlier work, \citep{ahmed_improved_2019} presented a variance-reduced Hill estimator, based on the difference between two Hill estimates of the source EVI: one computed on the $n$ coupled samples, and the other on the full set of $n+m$ available samples. Another related work is \citep{einmahl_variance-reduced_2024}, where variance-reduced estimators are presented for the Value-at-Risk, Expected Shortfall, and Expectile. The authors follow a similar method: correcting the target estimate with a difference between source-based estimates. Notably, their method is the only one that does not rely on a dependence model between target and source variables, which is often challenging to estimate when paired observations are limited.\\

This paper presents new variance-reduced EVI estimators in a semi-supervised framework, using the control variates method. The focus is placed on the Hill estimator, which can be formulated as a ratio of means, allowing the approximate control variates approach to be applied to both the numerator and denominator. The proposed transferred Hill estimator guarantees variance reduction and exhibits asymptotic independence of the relative variance reduction from the target and source EVI values under suitable conditions. The asymptotic relative variance reduction is also shown to be proportional to the tail dependence between target and source samples. \\

This paper is organized as follows. Section \ref{sec:EVT} presents central concepts of EVT and introduces the notations used throughout the article. It emphasizes the central role of the EVI and presents the considered EVI estimators. Section \ref{sec:CV} introduces the control variates method in its exact and approximate forms, for mean estimation and for ratio of means estimation. Section \ref{sec:appli_EVI} presents new variance-reduced EVI estimators using the approximate control variates method, and describes some properties of the proposed transferred Hill estimator. In Section \ref{sec:simu}, simulation results are presented, showcasing the efficiency of the proposed estimators and investigating the influence of key parameters on variance reduction. Finally, in Section \ref{sec:appli}, the new EVI estimators are applied to two practical cases: ice accretion on airplane wings, with spatial dependence, and water surge modeling, with multi-fidelity dependence.

\section{Extreme Value Theory}  \label{sec:EVT}
\subsection{Domains of attraction}
Extreme Value Theory (EVT) offers a statistical framework to analyze the tails of probability distributions, enabling the study of rare events that fall outside the range of observed data. The way extremes are defined varies across EVT approaches: the Block Maxima (BM) method or the Peaks-Over-Threshold (POT). In the BM framework, the sample of size $n$ is divided into $k$ sub-samples, each of size $r$, and only the maximum value within each block is retained. However, the BM method may fail to capture all extremes, as some blocks might contain multiple extremes while others may have none. The POT approach instead focuses on all data points that exceed a given high threshold $u$, such that $k$ extremes are considered. In this work, the POT framework is adopted, as it makes better use of the available information.

This section presents key results of EVT and introduces the EVI, the parameter of interest. For more details on EVT, see \citep{de_haan_extreme_2006} for a rigorous theoretical framework, and \citep{bousquet_extreme_2021} for a practical introduction.

Let $(Y_1, \dots, Y_n)$ be some i.i.d. samples of the random variable $Y\in \mathbb{R}$, and $Y_{1:n}\leq \dots\leq Y_{n:n}$ the associated order statistics. Let $F$ be their cumulative distribution function (cdf), with right endpoint $y^*=sup\{y: F(y)<1\}$. \\

A key result in EVT is the Fisher–Tippett–Gnedenko theorem stated below, which characterizes the asymptotic behavior of the sample maxima $Y_{n:n}$ and defines the max-domain of attraction.

\begin{theorem}[\citep{fisher_limiting_1928, gnedenko_sur_1943}] \label{th:GEV}
Assume there exist sequences of constants $(a_n > 0)$ and $(b_n)$ such that the normalized sample maximum $(Y_{n:n} - b_n)/a_n$ converges in distribution to a non-degenerate limit. Then 
\begin{align}\underset{n \to \infty}{\lim}\mathbb{P}\left(\frac{Y_{n:n}-b_n}{a_n} \leq y\right) =\underset{n \to \infty}{\lim} F^n(a_n y + b_n) = \text{GEV}_{\gamma}(y)\end{align}
where the limiting distribution is the Generalized Extreme Value (GEV) distribution, defined by
\begin{align}GEV_{\gamma}(y) := \begin{cases} \exp(-(1+\gamma y))^{-\frac{1}{\gamma}} \quad &\text{if } \gamma \neq 0,\quad  1+\gamma y>0; \\ \exp(-e^{-y}) \quad &\text{if } \gamma=0 ,\quad  y \in \mathbb{R}\end{cases}\end{align}
\end{theorem}

It can also be said that the distribution $F$ belongs to the max-domain of attraction $\mathcal{D}(GEV_\gamma)$ of a $GEV_\gamma$ distribution, where $\gamma$ is the EVI, characterizing the tail behavior of $F$. \\

The following theorem is of particular importance as it describes the asymptotic behaviour of the selected extremes in the POT framework, the values exceeding a predefined threshold.
\begin{theorem}[\citep{pickands_iii_statistical_1975, balkema_residual_1974}] \label{th:POT}
Let $F_u(y)=\mathbb{P}(Y-u\leq y|Y>u)$ be the distribution of exceedances over threshold $u$ and $\sigma(u)$ a positive scaling function. 
\begin{align}F \in \mathcal{D}(GEV_{\gamma}) \Leftrightarrow \underset{u \to y^*}{\text{lim}} F_u\left(\sigma(u) y\right) := GP_{\gamma}(y)\end{align}
where the limiting distribution is the Generalized Pareto Distribution (GPD), defined by \begin{align}GP_{\gamma}(y) = \begin{cases} 1 - (1+\gamma y)^{-\frac{1}{\gamma}} \quad &\text{if } \gamma \neq 0, \quad 0<y<(0\vee(-\gamma))^{-1}; \\ 1-exp(-y) \quad &\text{if } \gamma=0,\quad  y \in \mathbb{R} \end{cases}\end{align}
\end{theorem}
In other words, any distribution of data above a sufficiently high threshold can be approximated by a $GP_{\gamma}$ with EVI $\gamma$, provided that $F \in \mathcal{D}(GEV_{\gamma})$. This result provides the theoretical foundation for estimations in the POT framework.\\

With a positive EVI, the distribution has a heavy tail with an infinite right endpoint, meaning extreme events are more intense. It can model natural disasters \citep{bousquet_extreme_2021}, financial risks or insurance claims \citep{embrechts_modelling_1997}, or cyber attacks \citep{zhan_predicting_2015}. With a negative EVI, the distribution has a short tail with a finite right endpoint, making it suitable to model bounded quantities like lifespan \citep{einmahl_limits_2019} or physical constraints \citep{delaluz_estimating_2018}. When the EVI is zero, the distribution has a light tail with a finite or infinite right endpoint, meaning extreme events are possible but occur rarely, as in queuing times \citep{asmussen_extreme_1998}. These behaviors correspond respectively to the three classical domains of attraction: the Fréchet domain for $\gamma>0$, associated with heavy-tailed distributions such as the Pareto distribution; the Weibull domain for $\gamma<0$, associated with short-tailed distributions such as the Beta distribution; and the Gumbel domain for $\gamma=0$, associated with light-tailed distributions such as the Normal distribution.\\

Choosing the threshold $u$ in the POT framework is a delicate balance: setting it too low introduces bias by including non-extreme events, while setting it too high increases variance due to the reduced sample size. In practice, choosing $Y_{n-k:n}$ as threshold ensures a fixed number of exceedances $k$ rather than focusing on a set threshold value $u$. Using the random $(n-k)$-th order statistic as threshold results in EVI estimators with smaller variance than the ones with a deterministic threshold (remark 2.1 in \citep{bobbia_donsker_2025}). Threshold selection is still a key question in EVT; for further details, refer to \citep{caeiro_threshold_2015}. While the choice of $k$ affects the variance and the bias of EVI estimators, its optimization is not explored in this work.\\

$F$ being in the max-domain of attraction $F \in \mathcal{D}(GEV_{\gamma})$ can be expressed in terms of the asymptotic behavior of its tail quantile function $U$, as described in the first-order condition.
\begin{definition}[First-order extended regular variation \citep{de_haan_slow_1984}]
Let $a(\cdot)$ be a positive measurable function, and $U(\cdot)$ the tail quantile function defined for $t>1$ by $U(t)=F^{\leftarrow}\left(1-\frac{1}{t}\right)$, where $F^{\leftarrow}(y)=\text{inf}\{x:F(x)\geq y\}$ is the left-continuous inverse. Then \begin{align}F \in \mathcal{D}(GEV_{\gamma}) \Leftrightarrow \underset{t \to \infty}{\text{lim}} \frac{U(ty)-U(t)}{a(t)} = \begin{cases} \frac{y^{\gamma}-1}{\gamma} \quad &\text{if } \gamma \neq 0 ; \\ \text{ln }y \quad &\text{if } \gamma = 0\end{cases}\end{align}
$U(\cdot)$ is of first order extended regular variation, written $U \in \text{ERV}_{\gamma}$.
\end{definition}

The second-order condition quantifies the rate of convergence, which is essential for studying the finite-sample properties and convergence speed of EVI estimators \citep{de_haan_regular_1975}.
\begin{definition}[Second-order extended regular variation \citep{alves_note_2007}] \label{def:sec_order}
    Assume $A(\cdot)$ exists, a function such that $A(t)\underset{t \to \infty}{\longrightarrow}0$ and $\forall y > 0$,
    \begin{align}
        \underset{t \to \infty}{\text{lim}} \frac{\frac{U(ty) - U(t)}{a(t)} - \frac{y^\gamma - 1}{\gamma}}{A(t)} = H_{\gamma, \rho}(y) := \frac{1}{\rho} \left( \frac{y^{\gamma+\rho}-1}{\gamma + \rho} - \frac{y^\gamma-1}{\gamma} \right)
    \end{align}
    where $\rho \leq 0$ a parameter controlling the speed of convergence towards the limit law. Then $U(\cdot)$ is of second order extended regular variation, written $U \in 2\text{ERV}_{\gamma,\rho}$.
\end{definition}

\subsection{Extreme Value Index estimators} \label{sec:EVI_est}
The following section presents two commonly used EVI estimators that can be formulated as ratios of means, a necessary condition to apply the variance reduction method to be introduced in Section \ref{sec:CV}. For more details on EVI estimators, see Chapter 3 of \citep{de_haan_extreme_2006}.


\subsubsection{The Hill estimator}
The Hill estimator is a common estimator for a positive EVI.
\begin{definition}[Hill estimator \citep{hill_simple_1975}] \label{def:Hill}
Let $k \in \{1, \dots, n - 1\}$, and assume $Y_{n-k:n} > 0$, $F \in \mathcal{D}(GEV_{\gamma})$ and $\gamma>0$. The Hill estimator of the EVI $\gamma$ is defined as \begin{align}\widehat{\gamma}_H = \frac{1}{k} \sum_{i=1}^k \ln\left( \frac{Y_{n-i+1:n}}{Y_{n-k:n}} \right)\end{align}
\end{definition}

The Hill estimator is consistent under the first and second-order condition, provided that $\sqrt{k}A(\frac{n}{k}) \to 0$, with the number of extremes $k$ growing to infinity but remaining small compared to the total sample size $n$ (i.e. $k = k(n)\to \infty$ and $k/n \to 0$, as $n \to \infty$). The estimator's asymptotic behaviour can be characterized under the second-order condition of Definition \ref{def:sec_order}, as presented in Theorem 3.2.5 of \citep{de_haan_extreme_2006}. Its asymptotic bias is the limit of $\frac{A(n/k)}{1-\rho}$ where $\rho$ is the second-order parameter, and its asymptotic variance is equivalent to $\frac{\gamma^2}{k}$. If the second-order term $A(n/k)$ vanishes sufficiently fast as $k/n \to 0$, the estimator becomes asymptotically unbiased. The variance decreases as $k \to \infty$, highlighting the bias-variance trade-off in the choice of $k$. Bias-corrected EVI estimators have been proposed, one of which is presented below.

\subsubsection{The moment estimator}
\begin{definition}[Moment estimator \citep{dekkers_moment_1989}] \label{def:Moment}
Let $k \in \{1, \dots, n - 1\}$, and assume $Y_{n-k:n} > 0$, $F \in \mathcal{D}(GEV_{\gamma})$ and $\gamma > -1/2$. Define the empirical log-moments for $j \in \{1, 2\}$ by \begin{align}M^{(j)} = \frac{\frac{1}{n}\sum_{i=1}^n(\ln(Y_i) - \ln(Y_{n-k:n}))^j \1_{\{Y_i > Y_{n-k:n}\}}}{\frac{1}{n}\sum_{i=1}^n\1_{\{Y_i > Y_{n-k:n}\}}}\end{align} The moment estimator of the EVI $\gamma$ is defined as \begin{align}\widehat{\gamma}_{M} = M^{(1)} + 1 - \frac{1}{2} \left(1-\frac{\left(M^{(1)}\right)^2}{M^{(2)}}\right)^{-1}\end{align}
\end{definition}
The moment estimator is consistent under the first and second-order conditions. Compared to the Hill estimator, it exhibits lower bias and is not restricted to the heavy-tailed case.

\section{The control variates method} \label{sec:CV}
As discussed in the introduction, transfer learning in a semi-supervised framework is a promising approach for reducing the variance of EVI estimators. The transfer method explored in this work is based on control variates (CV), which are introduced in this section.

Let $(A_i)_{i=1 \dots n}$ be i.i.d. real samples of a random variable $A \in L^1$. The Monte Carlo estimator $\overline{A_n}$ of $\E[A]$ is defined as \begin{align}\overline{A_n}=\frac{1}{n}\sum_{i=1}^{n}A_i \label{eq:MC}\end{align}

To reduce the variance of the Monte Carlo estimator $\overline{A_n}$ of $\E[A]$, the idea of control variates is to consider an auxiliary variable $B$, correlated with the target random variable $A$. This approach is a form of statistical transfer learning as it leverages knowledge from a related variable $B$ to compensate for the lack of information about the target variable $A$. \\

Let $(A_i, B_i)_{i=1 \dots n}$ be i.i.d. real samples from the joint distribution of the random variables $A, B \in L^2$. Assume $\E[B]$ is known. The control variates estimator \citep{asmussen_variance-reduction_2007} of $\E[A]$ is defined as \begin{align}\widehat{A}_{CV} = \overline{A_n} + \alpha_c (\E[B] - \overline{B_n})\end{align} where the coefficient $\alpha_c$ is set to minimize the variance of $\widehat{A}_{CV}$: \begin{align}\alpha_c := \underset{\alpha \in \mathbb{R}}{\text{argmin}} \; \var\left(\overline{A_n} + \alpha (\E[B] - \overline{B_n})\right) = \frac{\cov(A, B)}{\var(B)} \label{eq:CV}\end{align}
The method can only reduce the variance compared to the baseline Monte Carlo estimator. The extent of variance reduction depends on the strength of the correlation between $A$ and $B$: the higher the absolute value of the correlation, the greater the reduction. When the correlation is zero, the control variates estimator coincides with the Monte Carlo one. Another strength of the method is that there is no added bias. A small plug-in bias may be added by the estimation of the coefficient $\alpha$ on the same sample as the one used to estimate $\E[A]$, but it is usually negligible \citep{owen_variance_2013}.\\

In practice, the assumption that the control variate mean $\E[B]$ is known is often unrealistic. The approximate control variates (ACV) method \citep{gorodetsky_generalized_2020} overcomes this by estimating the mean from additional data. Let $(A_i, B_i)_{i=1 \dots n}$ be i.i.d. real samples from the joint distribution of the random variables $A, B \in L^2$, and let $(B_i)_{i=n+1 \dots n+m}$ be additional i.i.d. samples of $B$. The approximate control variates estimator of $\E[A]$ is defined as \begin{align}\widehat{A}_{ACV} = \overline{A_n} + \alpha_c (\overline{B_{n+m}} - \overline{B_n})\end{align}

Assume the quantity of interest is a ratio of means $R=\E[A]/\E[C]$. The variance of its Monte Carlo estimator $R_{MC/MC}=\overline{A_n}/\overline{C_n}$ can be reduced by applying the control variates method to both numerator and denominator \citep{bocquet_control_2025}. The approximate control variates estimator for a ratio of means is defined as follows.
\begin{definition}[ACV/ACV estimator \citep{bocquet_control_2025}] \label{def:CV_CV}
Let $(A_i, B_i, C_i, D_i)_{i=1 \dots n}$ be i.i.d. real samples from the joint distribution of the random variables $A, B, C, D \in L^2$, where $B$ and $D$ are control variates. The ACV/ACV estimator of $R=\E[A]/\E[C]$ is defined as \begin{align}\widehat{R}_{\frac{ACV}{ACV}} = \frac{\overline{A_n} + \alpha (\overline{B_{n+m}} - \overline{B_n})}{\overline{C_n} + \beta (\overline{D_{n+m}} - \overline{D_n})} \end{align}
Assume $|\corr(B,D)|<1$. The optimal coefficients for the ACV/ACV estimator, guaranteeing variance reduction, are defined as \begin{alignat}{2}
\alpha = \frac{\var(D)\cov(A, B) - R \var(D)\cov(B, C)  + R \cov(B, D)\cov(C, D) -  \cov(B, D)\cov(A, D)}{\var(B)\var(D) - \cov(B,D)^2} \label{eq:alpha_opt} \\
\beta = \frac{ \cov(B, D)\cov(A, B) -  R \cov(B, D) \cov(B, C) + R \var(B)\cov(C, D) - \var(B)\cov(A, D)}{R \left(\var(B)\var(D) - \cov(B,D)^2 \right)} \label{eq:beta_opt} \end{alignat}
\end{definition}

\section{Variance-reduced Extreme Value Index estimators using control variates}\label{sec:appli_EVI}
The following section addresses the problem of variance reduction for EVI estimators that can be expressed as functions of ratios of means.  Let $Y^T, Y^S \in \mathbb{R}$ denote the target and source variables. A semi-supervised framework is considered: only a small number of paired target and source i.i.d. samples $(Y^T_i, Y^S_i)_{i=1\dots n}$ are available, supplemented by a larger set of additional source i.i.d. samples $(Y^S_i)_{i=n+1\dots n+m}$. This setting assumes that new samples cannot be acquired. The target-source relationship and the profusion of source data allow to improve the estimation of target data quantities. This is achieved through transfer learning with the control variates estimator presented in the previous section.

\subsection{Transferred Hill estimator}
The Hill estimator (Definition \ref{def:Hill}) is derived from the mean log-excess function, that converges to the EVI (Remark 1.2.3 of \citep{de_haan_extreme_2006}): \begin{align} \E[\ln(Y) - \ln(u) | Y > u] \underset{u \to y^*}{\longrightarrow} \gamma \end{align} Choosing the threshold $u$ as the $(n-k)$-th order statistic $Y_{n-k:n}$, and expressing the conditional expectation as a ratio, for $k = k(n)\to \infty$ and $k/n \to 0$ as $n \to \infty$, it gives: \begin{align} \frac{\E[(\ln(Y) - \ln(Y_{n-k:n})) \1_{{Y > Y_{n-k:n}}}]}{\E[\1_{{Y > Y_{n-k:n}}}]} \underset{n \to \infty}{\longrightarrow} \gamma \label{eq:ratio_Hill} \end{align} This leads to a formulation of the Hill estimator as a ratio of means: \begin{align}\widehat{\gamma}_{H} = \frac{\frac{1}{n}\sum_{i=1}^n(\ln(Y_i) - \ln(Y_{n-k:n})) \1_{\{Y_i > Y_{n-k:n}\}}}{\frac{1}{n}\sum_{i=1}^n\1_{\{Y_i > Y_{n-k:n}\}}} \label{eq:Hill_ratio}\end{align}

Using the notations of Section \ref{sec:CV}, the Hill estimator is expressed as \begin{align}\widehat{\gamma}^T_H = \frac{\overline{A_n}}{\overline{C_n}}\end{align} where \begin{align}A = (\ln(Y^T) - \ln(Y^T_{n-k:n})) \1_{\{Y^T > Y^T_{n-k:n}\}} \qquad C = \1_{\{Y^T > Y^T_{n-k:n}\}}\end{align} To be rigorous, $A$ and $C$ should be denoted $A_{k,n}$ and $C_{k,n}$ as they depend on the sample size $n$ and the number of extremes $k$, however we choose not to use the indices to keep the notations light.

This ratio of means form allows to apply the ACV/ACV estimator introduced in Definition \ref{def:CV_CV} to reduce variance, leading to the new transferred Hill estimator introduced hereafter.
\begin{definition}[Transferred Hill estimator]
Let $k \in \{1, \dots, n - 1\}$, and assume $Y^T_{n-k:n} > 0$, $F_T \in \mathcal{D}(GEV_{\gamma^T})$ and $\gamma^T>0$. Define the random variables \begin{align}A = (\ln(Y^T) - \ln(Y^T_{n-k:n})) \1_{\{Y^T > Y^T_{n-k:n}\}} \quad &B = (\ln(Y^S) - \ln(Y^S_{n-k:n})) \1_{\{Y^S > Y^S_{n-k:n}\}} \\ C = \1_{\{Y^T > Y^T_{n-k:n}\}} \quad &D = \1_{\{Y^S > Y^S_{n-k:n}\}}\end{align}The proposed transferred Hill estimator of the EVI, based on approximate control variates, is given by \begin{align}\widehat{\gamma}^{T}_{TH} = \frac{\overline{A_n} + \alpha (\overline{B_{n+m}} - \overline{B_n})}{\overline{C_n} + \beta (\overline{D_{n+m}} - \overline{D_n})}\end{align} where the coefficients $\alpha$ and $\beta$ are given by \begin{align}\alpha = \frac{\var(D)\cov(A, B) - R \var(D)\cov(B, C)  + R \cov(B, D)\cov(C, D) -  \cov(B, D)\cov(A, D)}{\var(B)\var(D) - \cov(B,D)^2} \end{align} \begin{align}\beta = \frac{ \cov(B, D)\cov(A, B) -  R \cov(B, D) \cov(B, C) + R \var(B)\cov(C, D) - \var(B)\cov(A, D)}{R \left(\var(B)\var(D) - \cov(B,D)^2 \right)} \end{align} where $R = \E[A]/\E[C]$.
\end{definition}

\begin{proposition}\label{prop_consistent}
    The estimator $\widehat{\gamma}^{T}_{TH}$ is consistent under the first and second conditions, as the Hill estimator, for $m \to \infty$, $k = k(n) \to \infty$, $k/n \to 0$ as $n \to \infty$.
\end{proposition}
Proof of Proposition \ref{prop_consistent} is given in Appendix \ref{sec:proof_consistent}.\\

In practice, the coefficients $\alpha$ and $\beta$ are estimated on the same dataset as the estimator $\widehat{\gamma}^{T}_{TH}$, which may introduce a plug-in bias, usually negligible as discussed in \citep{owen_variance_2013}. \\

The control variate is typically chosen to mirror the variable it controls. In this case, $B$ and $D$ are defined similarly to $A$ and $C$, using the source data in place of the target data. With this choice, the condition $|\corr(B,D)|<1$ is fulfilled and variance reduction is guaranteed. With the definitions of $A B, C, D$ given above, all four variables belong to $L^2$, ensuring the applicability of the control variates method.

The threshold for the control variates $B$ and $D$ is chosen to minimize the estimated variance of the estimator. The optimal source threshold was consistently found close to the ($n$-$k$)-th order statistic, as shown in Appendix \ref{sec:seuil}. Since estimating the variance for each possible threshold can be computationally expensive, and the benefit in variance reduction is minimal, the source threshold is set to $Y^S_{n-k:n}$. \\

As described in \citep{bocquet_control_2025}, it is delicate to determine which relations between the variables $A$, $B$, $C$, and $D$ allow for greater variance reduction. The potential for variance reduction can be assessed by examining the correlation between $C$ and $D$, which captures the co-occurrence of extremes in the target and source data. The correlation between $A$ and $B$ goes further, reflecting both the co-occurrence and the magnitude of the extremes, that is, how far above the threshold they go. Correlations $\corr(A,B)$ and $\corr(C,D)$ may offer some qualitative insight, but a linear relation with variance reduction cannot be established. In Section \ref{sec:dep}, it is shown that the control variates estimator works best in practice when the target and source extremes are correlated.

In essence, these correlations resemble the concept of tail dependence \citep{malevergne_investigating_2002}, which can also be used as an indicator of variance reduction. It measures the probability of observing a large value of $Y^T$ given that $Y^S$ is also large, or vice-versa, and is defined as: \begin{align} \lambda = \underset{u \to 1}{\lim} \, \mathbb{P}\left( Y^T > F_T^{-1}(u) \mid Y^S > F_S^{-1}(u) \right) = \underset{u \to 1}{\lim} \, \mathbb{P}\left( Y^S > F_S^{-1}(u) \mid Y^T > F_T^{-1}(u) \right) \end{align} It can be estimated by \begin{align} \widehat{\lambda} = \frac{1}{k} \sum_{i=1}^n \1\{Y^T > Y^T_{n-k:n}, Y^S > Y^S_{n-k:n}\}. \end{align}

In \citep{ahmed_extreme_2025}, tail dependence is a key indicator of variance reduction potential, as their estimator explicitly relies on an estimate of this quantity. In contrast, the estimator presented in this work implicitly benefits from source–target dependence without requiring its estimation. The following proposition establishes a connection between tail dependence and the relative variance reduction (RVR), a performance measure that quantifies the variance reduction achieved by a new estimator $\widehat{\gamma}_{\text{new}}$ compared to a baseline estimator $\widehat{\gamma}_{\text{base}}$: \begin{align}\text{RVR} = \frac{\text{Var}(\widehat{\gamma}_{\text{base}}) - \text{Var}(\widehat{\gamma}_{\text{new}})}{\text{Var}(\widehat{\gamma}_{\text{base}})}\end{align}
Here the baseline is the Hill estimator $\widehat{\gamma}^T_{H}$, compared to the new transferred Hill estimator $\widehat{\gamma}^T_{TH}$.

\begin{proposition} \label{prop} The asymptotic RVR of the transferred Hill estimator is studied for $k = k(n) \to \infty$, $k/n \to 0$ as $n \to \infty$.

\begin{enumerate}
\item  For any source distribution, the asymptotic RVR is 
\begin{flushleft}
\vspace{-25pt}
\hspace*{-80pt}
\begin{minipage}{1.4\textwidth}
    \begin{align}
        RVR \underset{n \to \infty}{\approx} \frac{m}{n(n+m)} \frac{1}{p^2} \frac{\cov(\epsilon_A',B)^2 \var(D) + \cov(\epsilon_A',D)^2 \var(B) - \cov(\epsilon_A',B)\cov(\epsilon_A',D)\cov(B,D)}{\var(B)\var(D)-\cov(B,D)^2}
    \end{align}
\end{minipage}
\end{flushleft}
\vspace{5pt}

\item For a heavy-tailed source distribution, i.e. $F^S \in \mathcal{D}(GEV_{\gamma^S})$ with $\gamma^S>0$, the scaled log-excesses $Z=\left( \frac{\ln(Y^S/u)}{\gamma^S} \bigg| Y^S>u\right)$ asymptotically follow a standard exponential distribution. The asymptotic RVR is 
    \begin{align}
        RVR \underset{n \to \infty}{\approx} \lambda^2 \frac{m}{n(n+m)} \frac{c_{AB}^2 + c_{AD}^2 \frac{2-p}{1-p} - c_{AB} c_{AD}}{p} \label{eq:asymp_RVR}
    \end{align}

where $p = \mathbb{P}(Y^T > Y^T_{n-k:n}) = \mathbb{P}(Y^S > Y^S_{n-k:n})$, \\$\epsilon_A' = \frac{A}{\gamma^T} - C \underset{n \to \infty}{\approx} (Z - 1) \1_{\{Y^T > Y^T_{n-k:n}\}}$,\\$c_{AD}$ and $c_{AB}$ defined as
    \begin{align*}
    c_{AD} &\underset{n \to \infty}{\approx} E\left[ Z - 1 \Bigg| Y^T > Y^T_{n-k:n}, Y^S > Y^S_{n-k:n}\right] \\
    c_{AB} &\underset{n \to \infty}{\approx} \E\left[ \left( Z - 1 \right) Z \Bigg| Y^T > Y^T_{n-k:n}, Y^S > Y^S_{n-k:n}\right]
    \end{align*}
\end{enumerate}
\end{proposition}

The proof is provided in Appendix \ref{sec:asymp_rvr}. This proposition establishes a clear link between RVR and squared tail dependence. The expression also introduces the coefficients $c_{AD}$ and $c_{AB}$, which do not correspond to any previously defined dependence measures. Notably, these coefficients are independent of both $\gamma^T$ and $\gamma^S$. \\

The implications of Proposition \ref{prop} are summarized in the following corollary.

\begin{corollary} \label{corollary}
The asymptotic RVR of the transferred Hill estimator is studied for $k = k(n) \to \infty$, $k/n \to 0$ as $n \to \infty$.

$\bullet$ For any source distribution, the asymptotic RVR is independent of the target EVI $\gamma^T$.

$\bullet$ For a heavy-tailed source distribution, i.e. $F^S \in \mathcal{D}(GEV_{\gamma^S})$ with $\gamma^S>0$, the asymptotic RVR is independent of the target and the source EVIs $\gamma^T$ and $\gamma^S$.
\end{corollary}
This highlights a key strength of the method: the variance reduction does not require similarity in tail heaviness between the source and target distributions.

\subsection{Transferred moment estimator}
Another variance-reduced EVI estimator can be derived from the moment estimator (Definition \ref{def:Moment}), by applying the ACV/ACV method to both moments as follows.
\begin{definition}[Transferred moment estimator]
Let $k \in \{1, \dots, n - 1\}$, and assume $Y^T_{n-k:n} > 0$, $F_T \in \mathcal{D}(GEV_{\gamma^T})$ and $\gamma^T \in \mathbb{R}$.
Define the random variables \begin{align}A = (\ln(Y^T) - \ln(Y^T_{n-k:n})) \1_{\{Y^T > Y^T_{n-k:n}\}} \qquad &B = (\ln(Y^S) - \ln(Y^S_{n-k:n})) \1_{\{Y^S > Y^S_{n-k:n}\}} \\ C = \1_{\{Y^T > Y^T_{n-k:n}\}} \qquad &D = \1_{\{Y^S > Y^S_{n-k:n}\}} \\ G = \left(\ln(Y^T) - \ln(Y^T_{n-k:n})\right)^2 \1_{\{Y^T > Y^T_{n-k:n}\}} \qquad &H = \left(\ln(Y^S) - \ln(Y^S_{n-k:n})\right)^2 \1_{\{Y^S > Y^S_{n-k:n}\}}\end{align}
The variance-reduced empirical log-moments for $j \in \{1, 2\}$ are defined by \begin{align}M^{(1)}_{CV} = \frac{\overline{A_n} + \alpha (\overline{B_{n+m}} - \overline{B_n})}{\overline{C_n} + \beta (\overline{D_{n+m}} - \overline{D_n})} \qquad M^{(2)}_{CV} = \frac{\overline{G_n} + \alpha' (\overline{H_{n+m}} - \overline{H_n})}{\overline{C_n} + \beta' (\overline{D_{n+m}} - \overline{D_n})}\end{align}
The transferred moment estimator of the EVI, using control variates, is given by \begin{align} \widehat{\gamma}^{T}_{TM} = M^{(1)}_{CV} + 1 - \frac{1}{2} \left(1-\frac{\left(M^{(1)}_{CV}\right)^2}{M^{(2)}_{CV}}\right)^{-1} \end{align}
where the coefficients $(\alpha, \beta, \alpha', \beta')$ are chosen to minimize the variance of each empirical moment and given by 
\begin{align}\alpha = \frac{\var(D)\cov(A, B) - M^{(1)} \var(D)\cov(B, C)  + M^{(1)} \cov(B, D)\cov(C, D) -  \cov(B, D)\cov(A, D)}{\var(B)\var(D) - \cov(B,D)^2} \end{align}
\begin{align}\beta = \frac{\cov(B, D)\cov(A, B) -  M^{(1)}\cov(B, D) \cov(B, C) + M^{(1)}\var(B)\cov(C, D) - \var(B)\cov(A, D)}{M^{(1)} \left(\var(B)\var(D) - \cov(B,D)^2 \right)}. \end{align}
\begin{align}\alpha' = \frac{\var(D)\cov(G, H) - M^{(2)} \var(D)\cov(H, C)  + M^{(2)} \cov(H, D)\cov(C, D) -  \cov(H, D)\cov(G, D)}{\var(H)\var(D) - \cov(H,D)^2} \end{align}
\begin{align}\beta' = \frac{\cov(H, D)\cov(G, H) -  M^{(2)}\cov(H, D) \cov(H, C) + M^{(2)}\var(H)\cov(C, D) - \var(H)\cov(G, D)}{M^{(2)} \left(\var(H)\var(D) - \cov(H,D)^2\right)} \end{align} where $M^{(1)} = \E[A]/\E[C]$ and $M^{(2)} = \E[G]/\E[C]$.
\end{definition}

The estimator $\widehat{\gamma}^{T}_{TM}$, as the moment estimator, is consistent for $k = k(n) \to \infty$, $k/n \to 0$ as $n \to \infty$.

With the definitions of $A, B, C, D, G, H$ given above, all six variables belong to $L^2$, ensuring the applicability of the control variates method. As the variables fulfill the conditions $|\corr(B,D)|<1$ and $|\corr(H,D)|<1$, the variance reduction is guaranteed for each moment individually. However, the expressions of the coefficients $(\alpha, \beta, \alpha', \beta')$ proposed here are not optimal: they minimize the variance of each moment $M^{(1)}_{CV}$ and $M^{(2)}_{CV}$ independently, and not the variance of the EVI estimator $\widehat{\gamma}^{T}_{TM}$. This can lead to a variance increase. Determining the optimal coefficients is an open question to be addressed. Here, the goal is simply to demonstrate that the method applies to another EVI estimator.

\section{Simulation study} \label{sec:simu}
Various results are presented here in order to study the properties of the variance-reduced EVI estimators introduced in this work. Reported results are averaged over 10,000 independent repetitions. Performance is measured with the RVR. Throughout all experiments, the number of extremes is set to $k = 0.1n$.\\

The target samples $(Y^T_i)_{i=1\dots n}$ and source samples $(Y^S_i)_{i=1\dots n+m}$ are simulated with Pareto marginals, defined by the cdf \begin{align}F(y) = \begin{cases} 1 - \left(\frac{y}{y_m}\right)^{-\gamma} \quad &\text{if }y \geq y_m>0 \\ 0 \quad &\text{otherwise}    
\end{cases}\end{align} where $y_m>0$ is the scale parameter and $\gamma>0$ is the EVI, set to $\gamma^T$ and $\gamma^S$ for the target and source samples respectively. This distribution is selected because it satisfies the assumptions under which the studied EVI estimators become asymptotically unbiased. It also allows direct control of the EVI value. The scale parameter is set to $y_m = 10^{-3}$ to ensure all samples are strictly positive, as required by the EVI estimators considered in this study.

The dependence between the source and target samples is modeled with a Gumbel copula $C_\mathcal{G}$, which is a bivariate distribution function with uniform marginals that models strong dependence in the upper tail. It is defined as \begin{align}C_\mathcal{G} : (u_1,u_2)\in[0,1]^2 \;\mapsto\; exp\Big(-\big((-log(u_1))^\theta + (-log(u_2))^\theta)\big)^{\frac{1}{\theta}}\Big)\end{align} where $\theta \geq 1$ controls how strong the the tail dependence is.

\subsection{Comparison with the baseline estimators}
In this section, the proposed variance-reduced estimators, the transferred Hill and the transferred moment estimators, are compared to their respective baselines. The MLE-based estimator from \citep{ahmed_extreme_2025} is also included for comparison, with the parameter $g$ set to the target EVI as recommended for optimal results. It is important to note that the variance reduction approaches start from very different baselines: the Hill estimator exhibits significantly lower variance than the MLE estimator, but also a higher bias. The three different variance-reduced estimators and their baselines are presented together to give an overview of possible results, but not for comparison. The Semi-Supervised Estimator (SSE) from \citep{ahmed_extreme_2025} is thus only included for context, since it is the only method with available code employing a semi-supervised framework for a variance-reduced EVI estimator.

\begin{figure}[!ht]
    \centering
    \includegraphics[width=\linewidth]{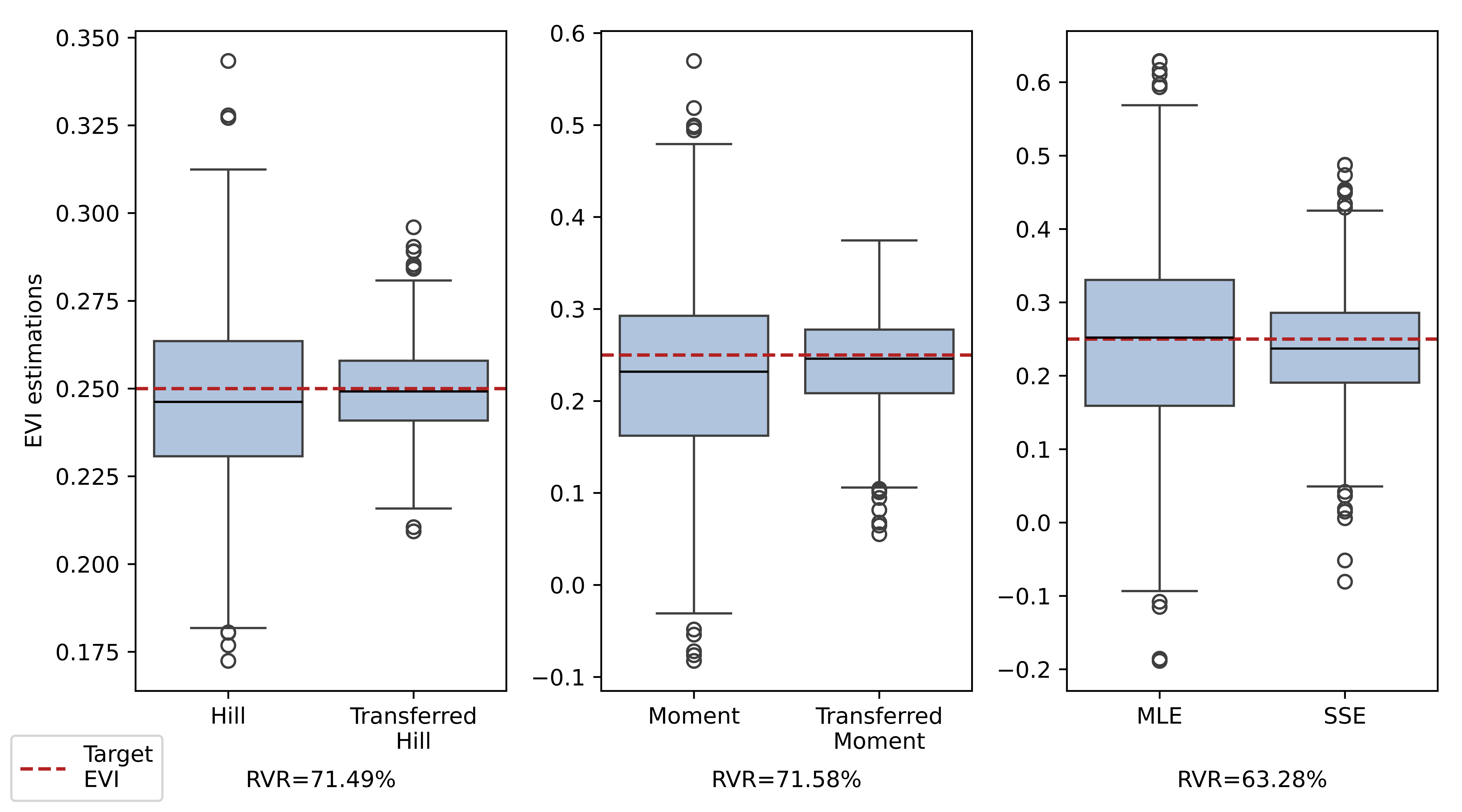}
    \captionsetup{justification=centering}
    \caption{Boxplots of EVI estimations for different methods with strong target-source dependence \\
    \textit{($\gamma^T=0.25$, $\gamma^S=0.5$, $n=1,000$, $k=100$, $m=5,000$, $\theta=10$)}}
    \label{fig:boxplots_methods_10}
\end{figure}

In Figure \ref{fig:boxplots_methods_10}, the Gumbel copula parameter is set to $\theta=10$, resulting in a highly favorable setting with strong dependence between target and source samples: $\widehat{\corr}(A,B)=0.99$, $\widehat{\corr}(C,D)=0.92$, and $\widehat{\lambda}=0.93$. In this setting, the transferred estimators achieve an RVR of $71\%$ when respectively compared to the Hill and moment baseline estimators.
 
\begin{figure}[!ht]
    \centering
    \includegraphics[width=\linewidth]{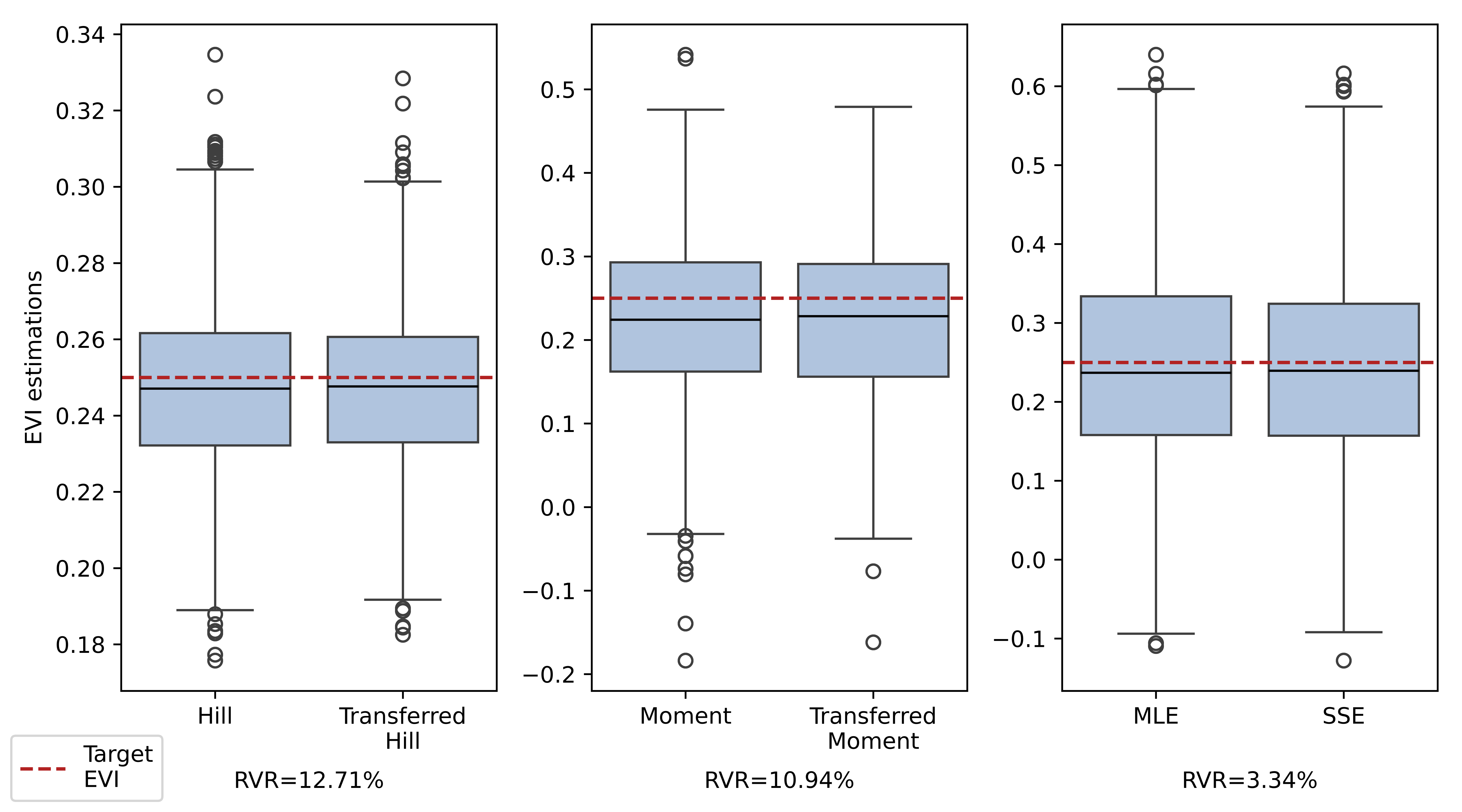}
    \captionsetup{justification=centering}
    \caption{Boxplots of EVI estimations for different methods with weaker target-source dependence \\
    \textit{($\gamma^T=0.25$, $\gamma^S=0.5$, $n=1,000$, $k=100$, $m=5,000$, $\theta=1.4$)}}
    \label{fig:boxplots_methods_1,4}
\end{figure}

In Figure \ref{fig:boxplots_methods_1,4}, the Gumbel copula parameter is set to $\theta=1.4$, resulting in a less favorable setting with only mild dependence between the target and source data: $\widehat{\corr}(A,B) = 0.50$, $\widehat{\corr}(C,D) = 0.35$, and $\widehat{\lambda} = 0.41$. The RVR achieved by the transferred Hill and moment estimators remains similar, $12\%$ and $10\%$ respectively, but it is notably lower than in the stronger dependence scenario. The impact of dependence on variance reduction is analysed in greater detail in Section \ref{sec:dep}.

The transferred estimators exhibit a slight bias compared to their baselines; this bias comes from the estimation of the control variates coefficients, however it is often negligible in practice as discussed in \citep{owen_variance_2013}.

\subsection{Influence of the dependence between target and source samples on the variance reduction} \label{sec:dep}

\begin{figure}[!ht]
    \begin{subfigure}[b]{0.49\linewidth}
        \centering
        \includegraphics[width=\linewidth]{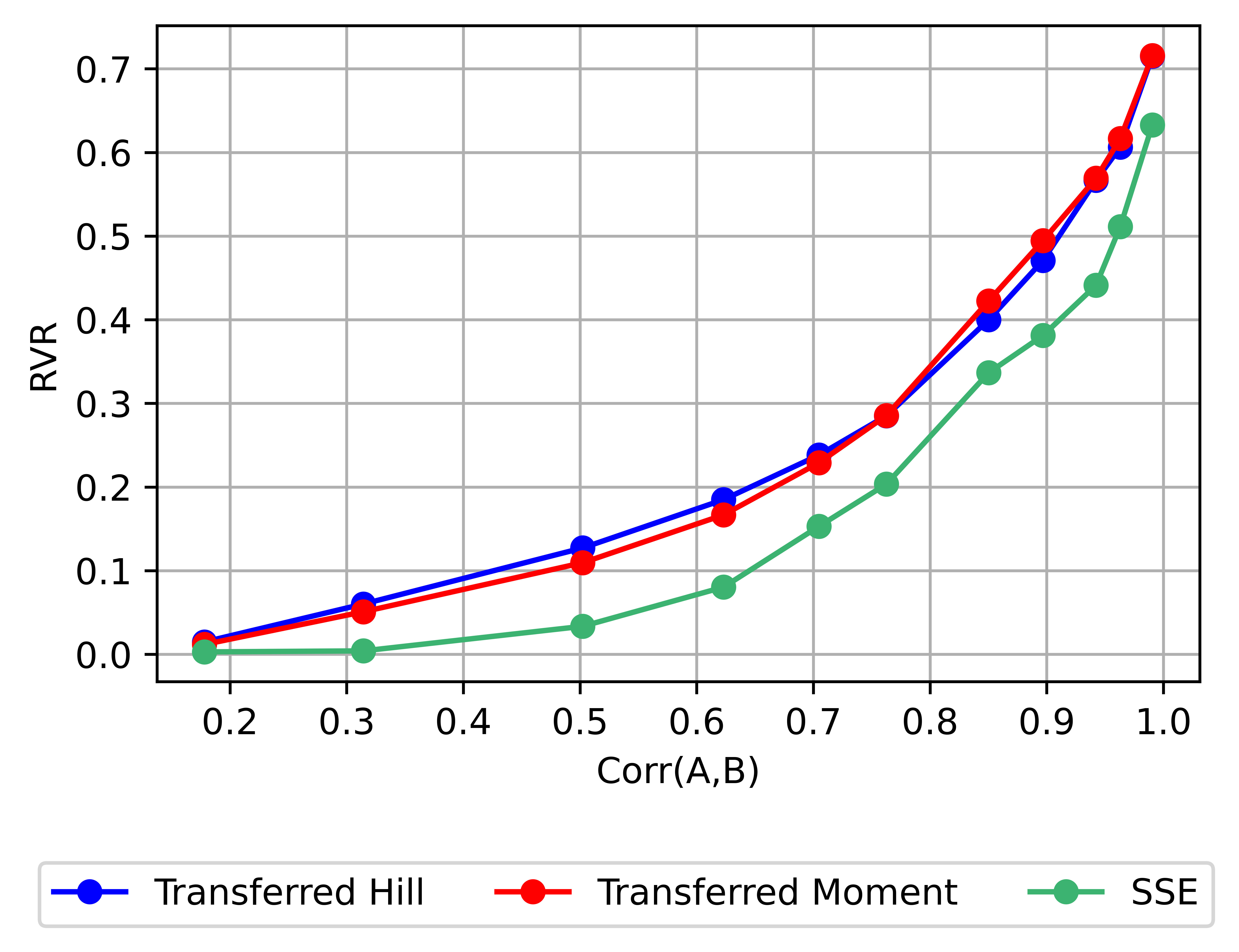}
        \subcaption{RVR vs tail dependence $\widehat{\lambda}$}
    \end{subfigure}
    \hfill
    \begin{subfigure}[b]{0.49\linewidth}
        \centering
        \includegraphics[width=\linewidth]{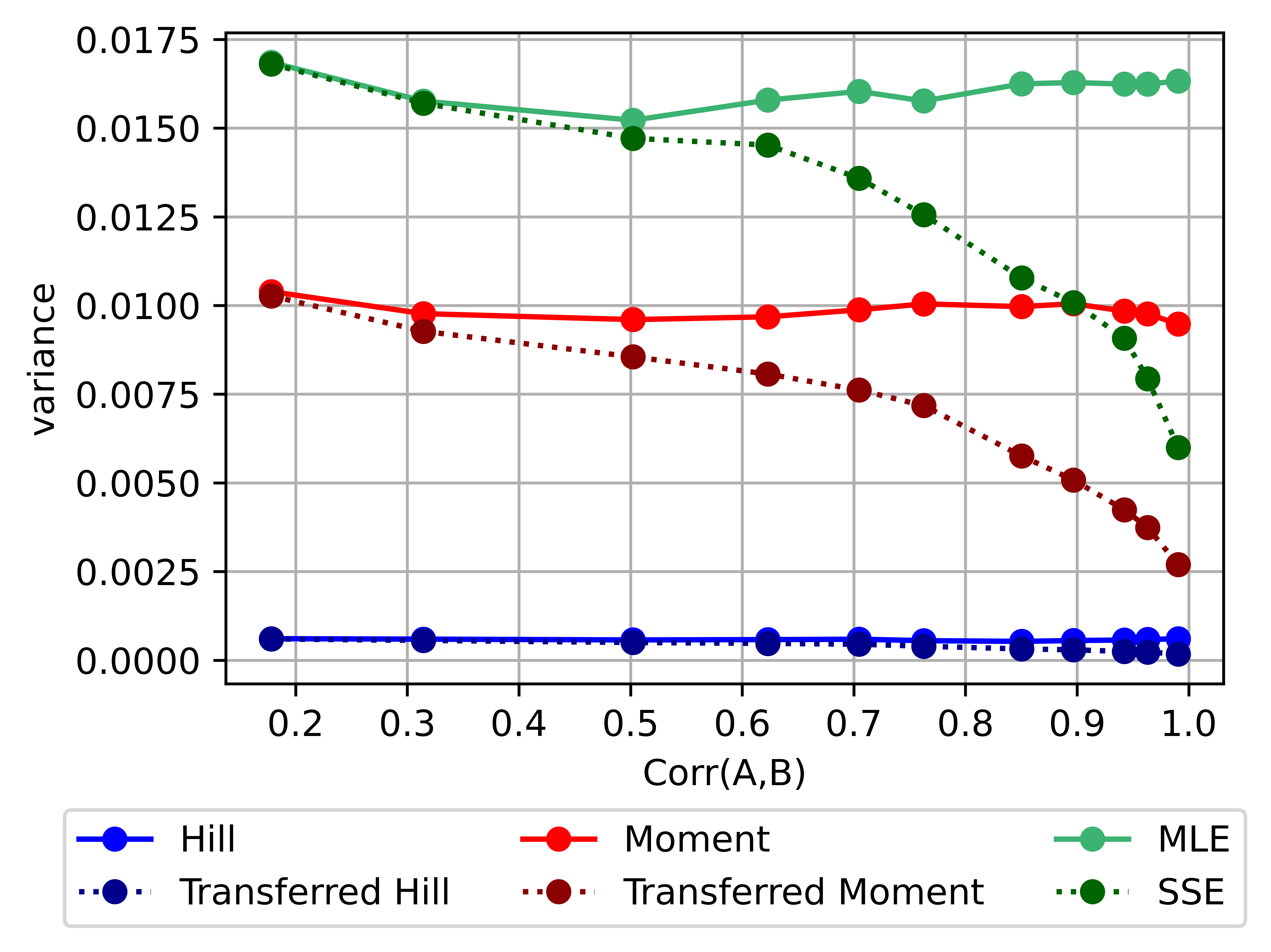}
        \subcaption{Variance vs tail dependence $\widehat{\lambda}$}
    \end{subfigure}
    \captionsetup{justification=centering}
    \caption{Variance reduction according to the target-source tail dependence\\\textit{($\gamma^T=0.25$, $\gamma^S=0.5$, $n=1,000$, $k=100$, $m=5,000$)}}
    \label{fig:rvr_dep}
\end{figure}

Figure \ref{fig:rvr_dep} presents the variance reduction achieved by the transferred Hill and moment estimators across various target-source dependence settings, with the SSE estimator from \citep{ahmed_extreme_2025} included for comparison. The results illustrate that stronger dependence between target and source increases variance reduction. In particular, the figure highlights the theoretical link between asymptotic RVR and tail dependence for the transferred Hill estimator, as established in Proposition \ref{prop}.

\subsection{Influence of the parameters on the variance reduction}
In this section, the Gumbel copula parameter is set to $\theta=5$, creating a favorable framework where $\widehat{\corr}(A,B) = 0.96$, $\widehat{\corr}(C,D) = 0.84$, and $\widehat{\lambda} = 0.86$. \\

\begin{figure}[!ht]
    \centering
    \begin{subfigure}[b]{0.49\linewidth}
        \includegraphics[width=\linewidth]{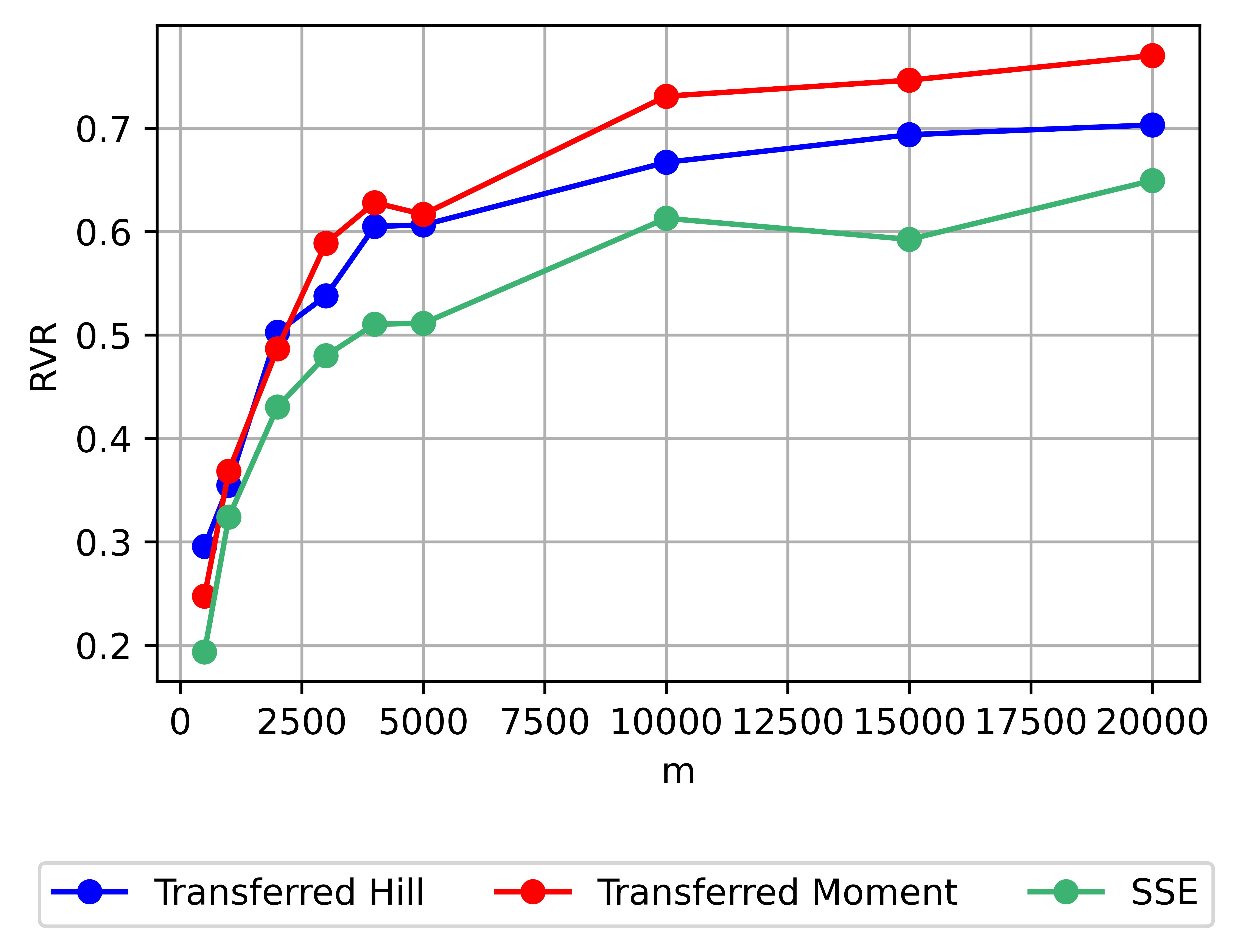}
        \subcaption{Variance vs number of additional source samples $m$}
    \end{subfigure}
    \hfill
    \begin{subfigure}[b]{0.49\linewidth}
        \includegraphics[width=\linewidth]{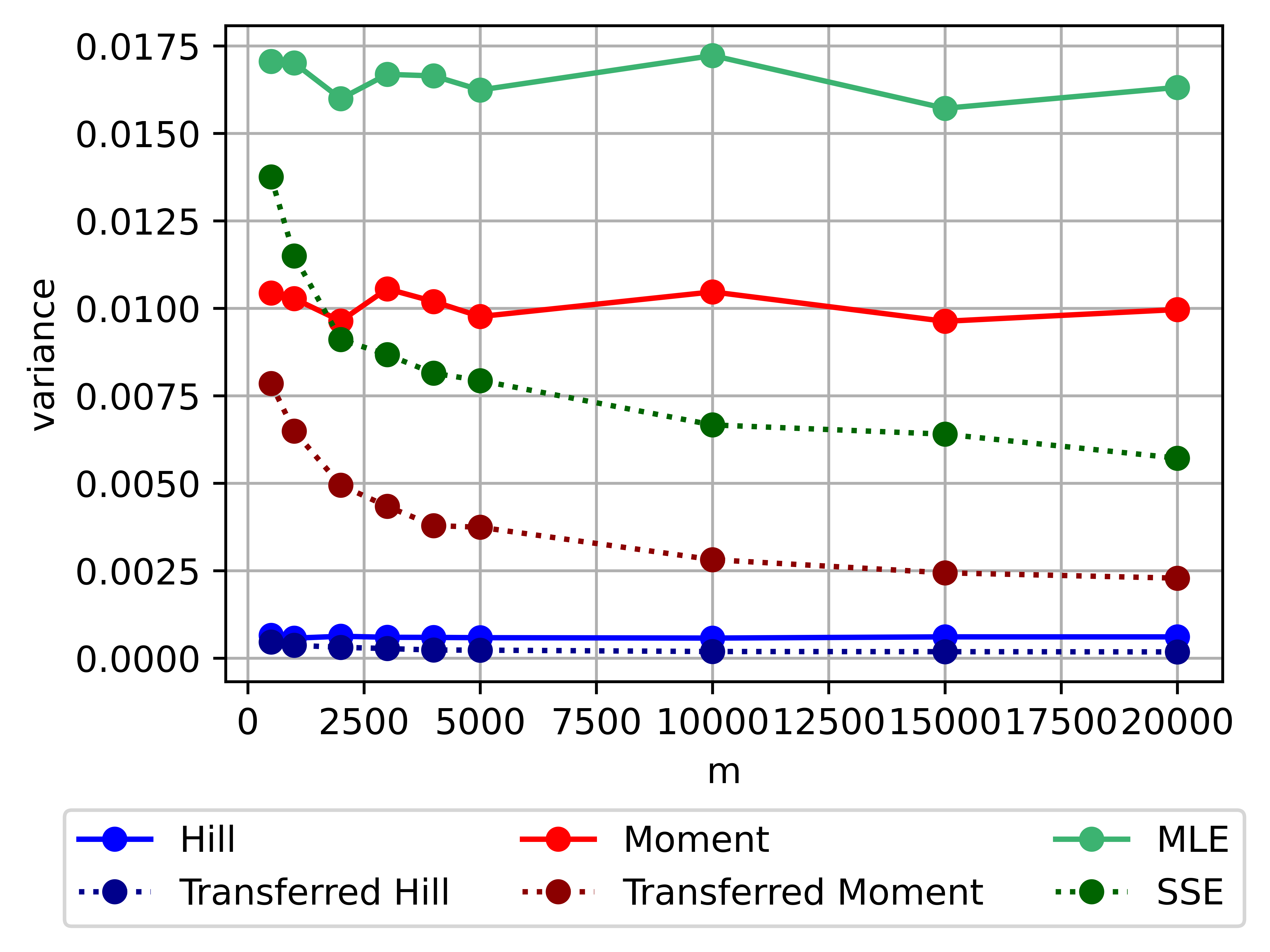}
        \subcaption{RVR vs number of additional source samples $m$}
    \end{subfigure}
    \caption{\textit{($\gamma^T=0.25$, $\gamma^S=0.5$, $n=1,000$, $k=100$, $\theta=5$)}}
    \label{fig:m}
\end{figure}

The first parameter studied is $m$, the number of additional source samples. Figure \ref{fig:m} shows that increasing the number of additional source samples leads to greater variance reduction. This result is expected, as the approximate control variates method benefits from a more accurate estimation of the control variate mean, which improves with larger $m$. Indeed, when $m \to \infty$, the variance reduction obtained with approximate control variates approaches that of the exact control variates method.\\

\begin{figure}[!ht]
    \centering
    \begin{subfigure}[b]{0.49\linewidth}
        \includegraphics[width=\linewidth]{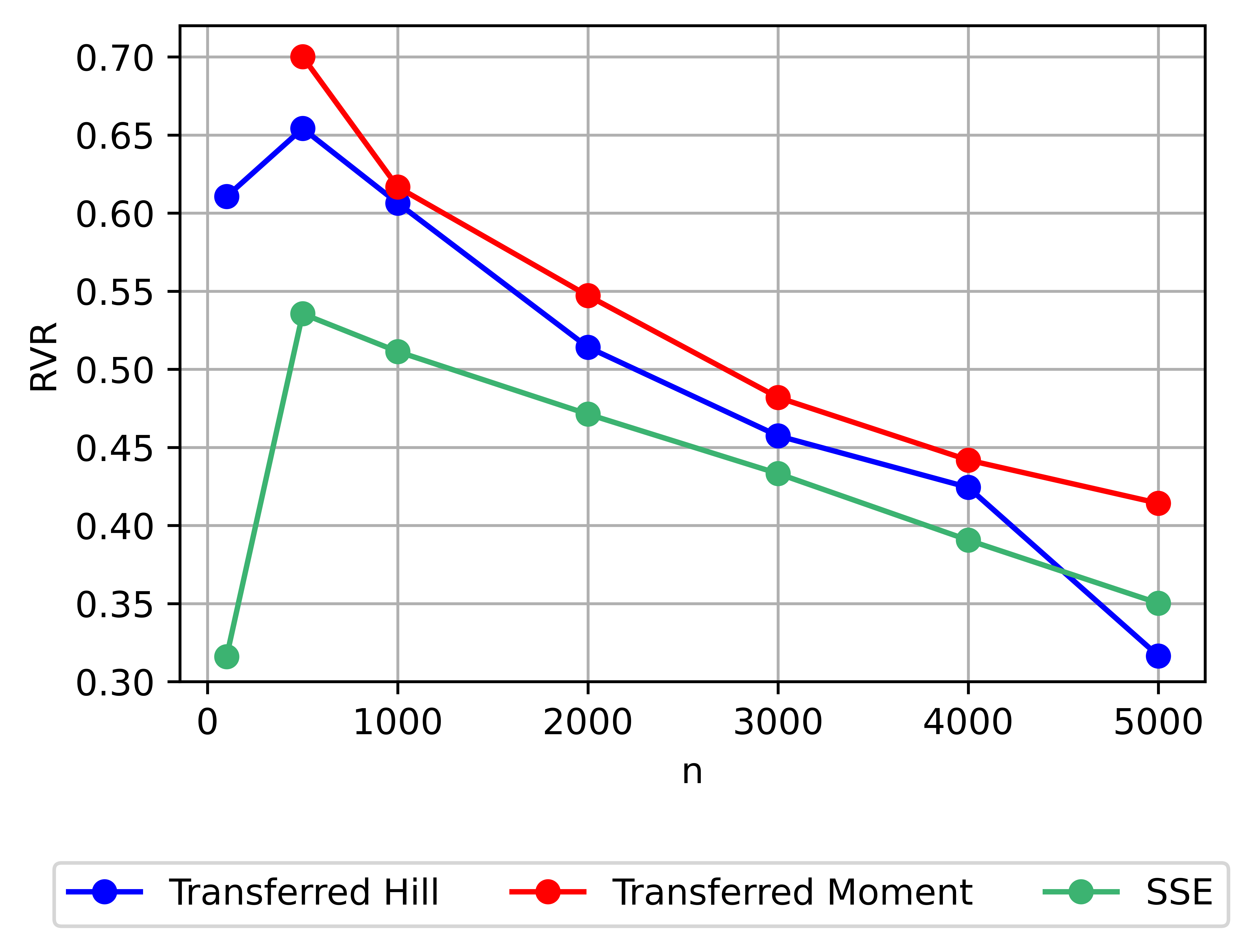}
        \subcaption{Variance vs number of coupled samples $n$}
    \end{subfigure}
    \hfill
    \begin{subfigure}[b]{0.49\linewidth}
        \includegraphics[width=\linewidth]{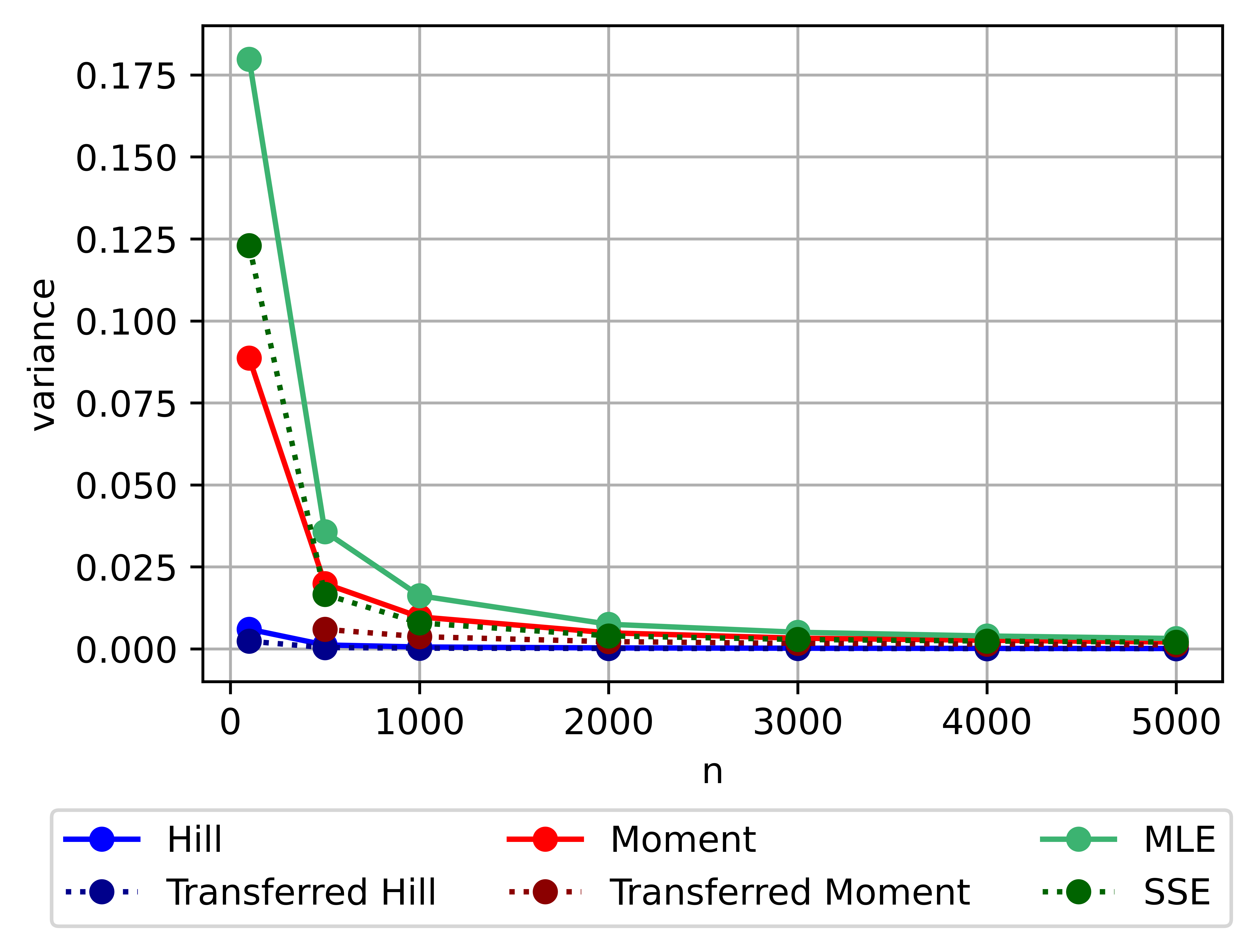}
        \subcaption{RVR vs number of coupled samples $n$}
    \end{subfigure}
    \caption{\textit{($\gamma^T=0.25$, $\gamma^S=0.5$, $m=5,000$, $\theta=5$)}}
    \label{fig:n}
\end{figure}
The parameters $n$ and $k$ are studied next, where $n$ denotes the number of coupled samples and $k$ the number of extremes considered. For the sake of simplicity, only the value of $n$ is described in the figure, setting $k = 0.1n$. As shown in Figure \ref{fig:n}, smaller sample sizes $n$ lead to greater variance reduction, which is expected given the higher variance of the baseline estimators in these scenarios. Notably, for $n = 100$ and $k = 10$, the variance of the Hill estimator is significantly reduced using the transferred Hill estimator. This is not the case for the moment estimator, which is excluded from the figure due to instability. This likely stems from the fact that the coefficients employed are optimized for each moment individually, rather than for reducing the overall variance of the EVI estimator. Additionally, they are more challenging to estimate accurately, leading to the presence of outliers and increased variability. In the small-sample scenario with $n = 100$ and $k = 10$, the SSE estimator from \citep{ahmed_extreme_2025} also underperforms.\\

\begin{figure}[!ht]
    \centering
    \begin{subfigure}[b]{0.49\linewidth}
        \includegraphics[width=\linewidth]{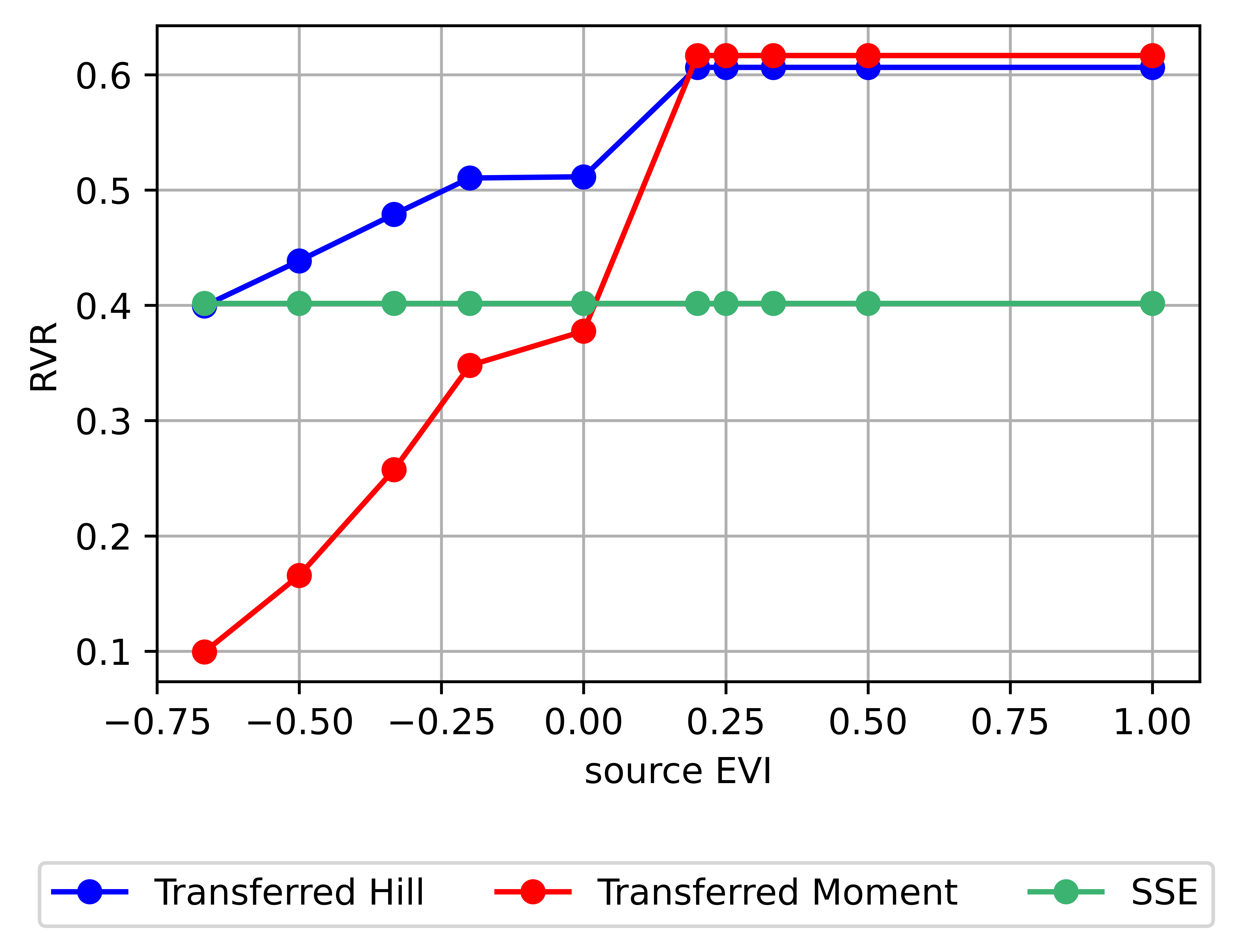}
        \subcaption{Variance vs source EVI $\gamma^{S}$}
    \end{subfigure}
    \hfill
    \begin{subfigure}[b]{0.49\linewidth}
        \includegraphics[width=\linewidth]{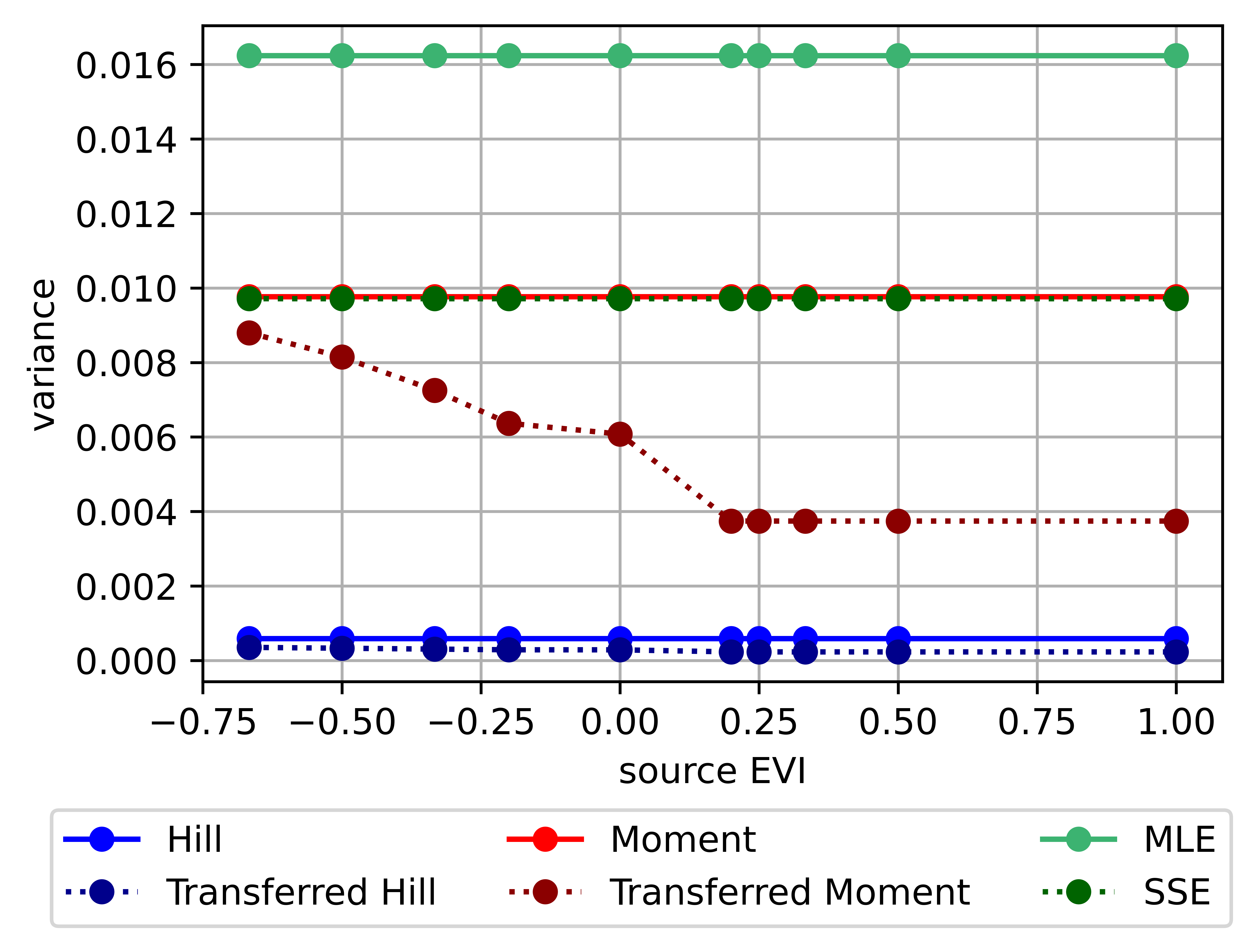}
        \subcaption{RVR vs source EVI $\gamma^{S}$}
    \end{subfigure}
    \caption{\textit{($\gamma^T=0.5$, $n=1,000$, $k=100$, $m=5,000$, $\theta=5$)}}
    \label{fig:source_EVI}
\end{figure}
The influence of different source EVI values is illustrated in Figure \ref{fig:source_EVI}. The values of $\gamma^S$ tested are $\{-0.66, -0.5, -0.33, -0.2, 0, 0.2, 0.25, 0.33, 0.5, 1\}$. Source samples with $\gamma^S = 0$ are simulated with a standard Normal marginal, while those with $\gamma^S<0$ follow a Beta marginal.

The results show that the variance reduction achieved by the transferred Hill estimator remains constant for a range of positive source EVI values, as described in Corollary \ref{corollary}. When the source EVI is non-positive, the variance reduction is lower but still satisfactory. The same results are observed with the transferred moment estimator. In contrast, the SSE estimator from \citep{ahmed_extreme_2025} maintains stable performance across all source EVI values. This is because the source samples are transformed to have a fixed, chosen EVI $g$, making the initial source EVI irrelevant to the variance reduction.\\

\begin{figure}[!ht]
    \centering
    \begin{subfigure}[b]{0.49\linewidth}
        \includegraphics[width=\linewidth]{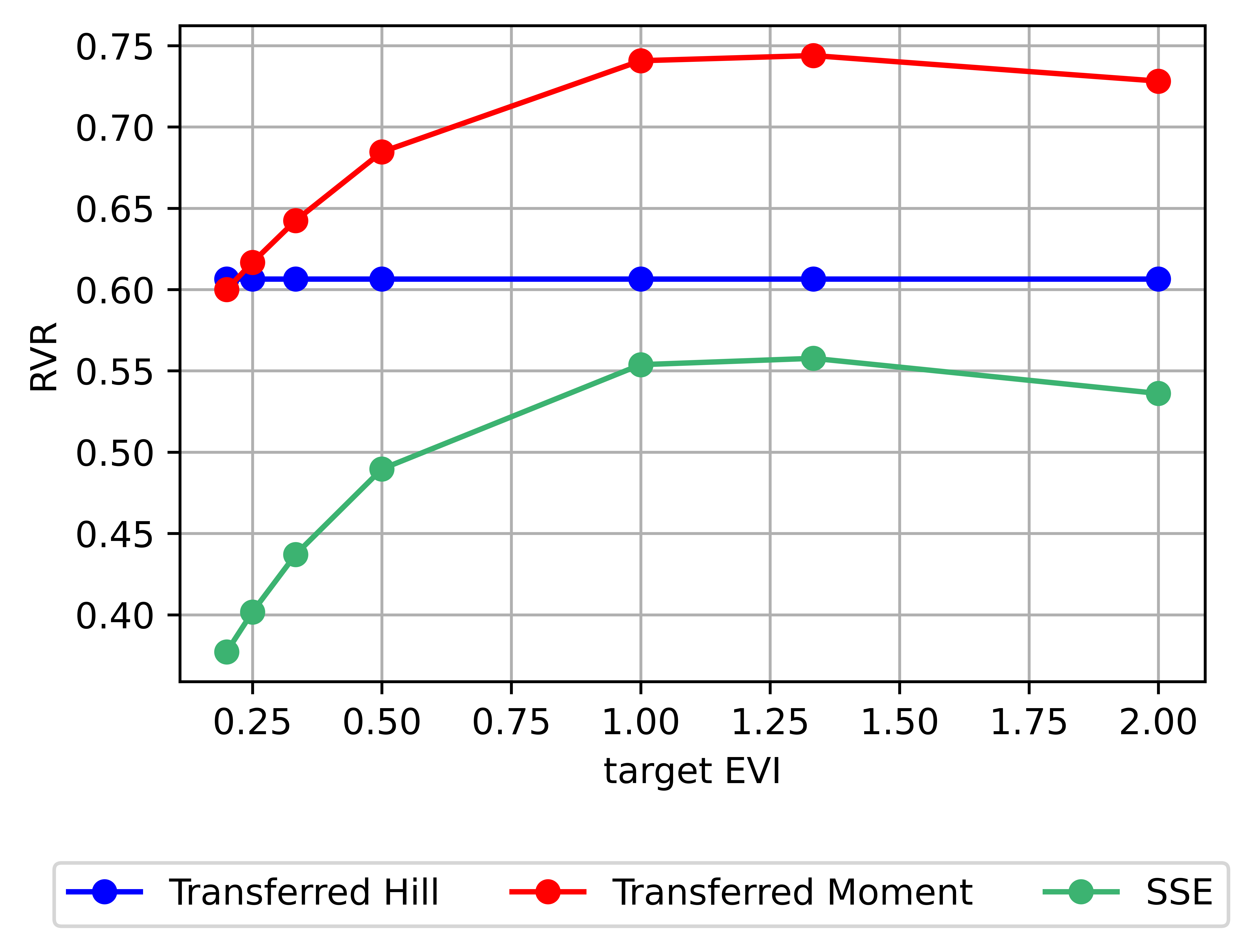}
        \subcaption{Variance vs target EVI $\gamma^{T}$}
    \end{subfigure}
    \hfill
    \begin{subfigure}[b]{0.49\linewidth}
        \includegraphics[width=\linewidth]{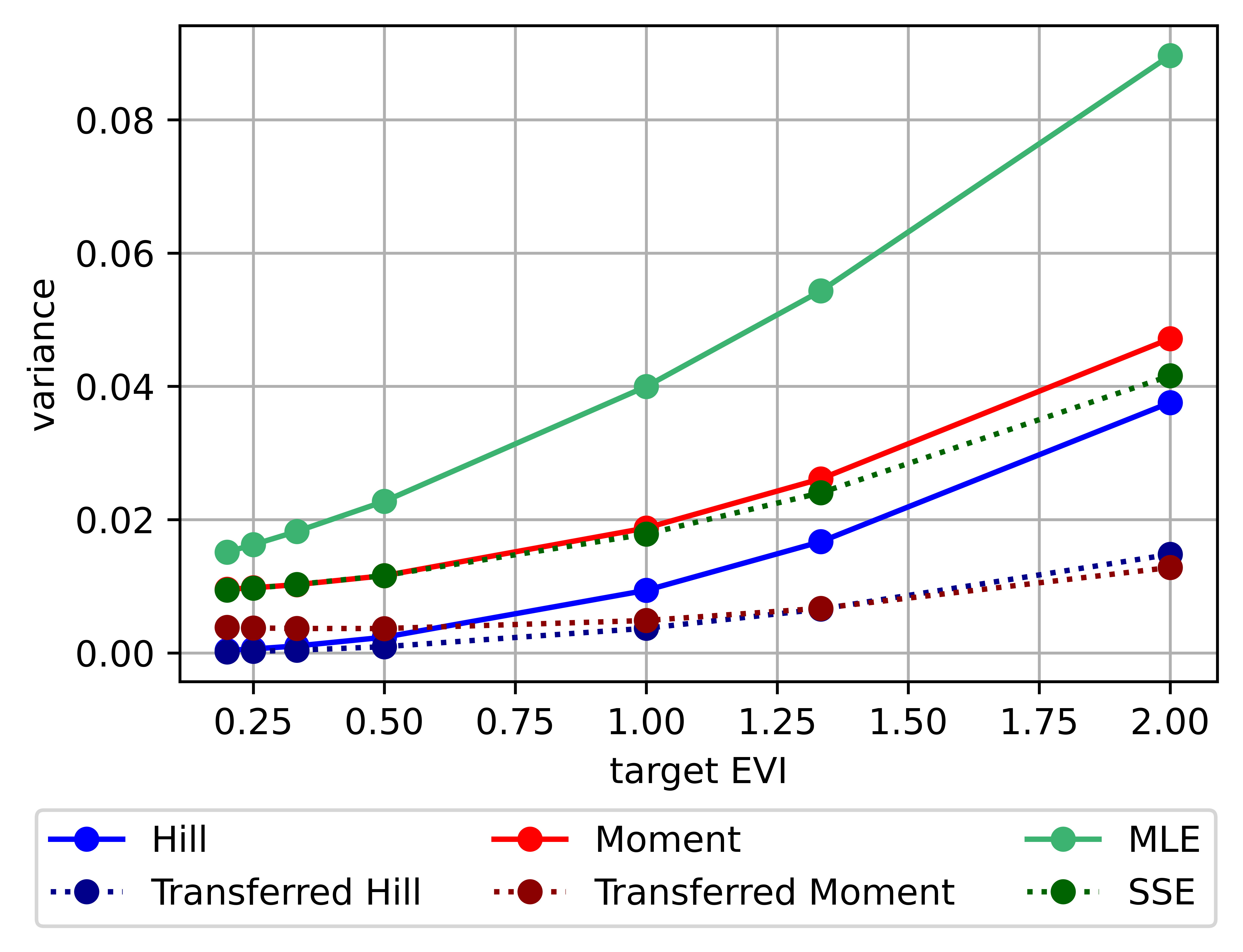}
        \subcaption{RVR vs target EVI $\gamma^{T}$}
    \end{subfigure}
    \caption{\textit{($\gamma^S=0.5$, $n=1,000$, $k=100$, $m=5,000$, $\theta=5$)}}
    \label{fig:EVI}
\end{figure}
The variance reduction for different target EVI values is also studied and illustrated in Figure \ref{fig:EVI}. The values of $\gamma^T$ tested are $\{ 0.2, 0.25, 0.33, 0.5, 1, 1.33, 2\}$. The variance reduction achieved with the transferred Hill estimator is independent of the target EVI estimated, as shown in Corollary \ref{corollary}.

\section{Application} \label{sec:appli}
In this section, two applications are presented. The first one is based on a multi-fidelity model of water surge, where the high-fidelity (HF) output is the target sample, and the low-fidelity (LF) output is the source. Since both HF and LF models approximate the same physical process, their outputs are inherently dependent. The second application uses a physics-based model of ice accretion on airplane wings. Two output variables, computed at different locations along the same wing profile, are considered as the source and target samples. The dependence between them is due to spatial continuity in the underlying physical phenomenon. These two different examples highlight that the proposed variance reduction method for EVI estimators can be applied in a wide range of scenarios.

The results presented in the following are obtained by bootstrapping 500 random subsamples of size $n$ from the $n+m$ available joint observations, to assess the statistical variability and robustness of the methods tested.

\subsection{Water surge model}
The water surge model is an analytical physics-based model initially developed in \citep{baudin_openturns_2017} for flood risk assessment. The high-fidelity (HF) version is derived from a simplified form of the one-dimensional Saint-Venant equations, under the assumptions of a uniform, constant flow rate and large rectangular cross-sections. The low-fidelity (LF) model used here was introduced in \citep{espoeys_optimisation_2025}.

\begin{figure}[!ht]
    \centering
    \includegraphics[width=0.8\linewidth]{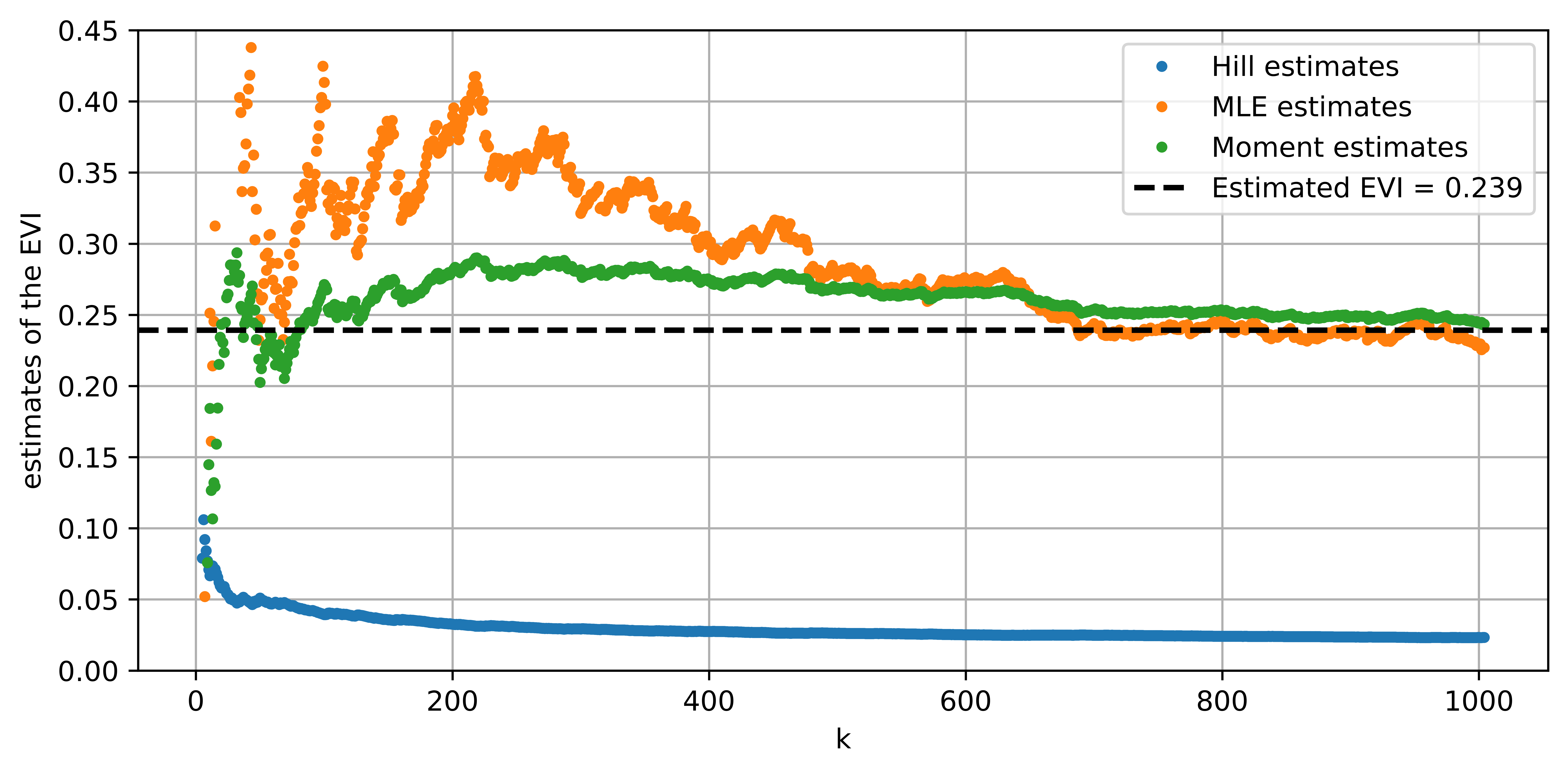}
    \caption{Hill plot on 10,000 water surge HF samples}
    \label{fig:flood_evi}
\end{figure}

A reference value for the target EVI is obtained by computing the Hill, moment, and maximum-likelihood estimators on a large HF dataset of size 10,000. Because the HF model is available and fast to evaluate, generating such a large sample is possible and yields a highly accurate EVI estimate. The Hill plot method \citep{caeiro_threshold_2015} allows to study the stability of the EVI estimates across values of $k$, and allows here to check that $k=0.1n$ is a good choice. Figure \ref{fig:flood_evi} confirms that $k=1,000$ lies in a stable region and the target EVI is estimated to be 0.239 with the MLE, known to be asymptotically unbiased. The figure also highlights that the Hill estimator is biased for the water surge dataset, as discussed theoretically in Section \ref{sec:EVI_est}.

\begin{figure}[!ht]
    \centering
    \begin{subfigure}[b]{0.49\linewidth}
        \centering
        \includegraphics[height=0.25\textheight]{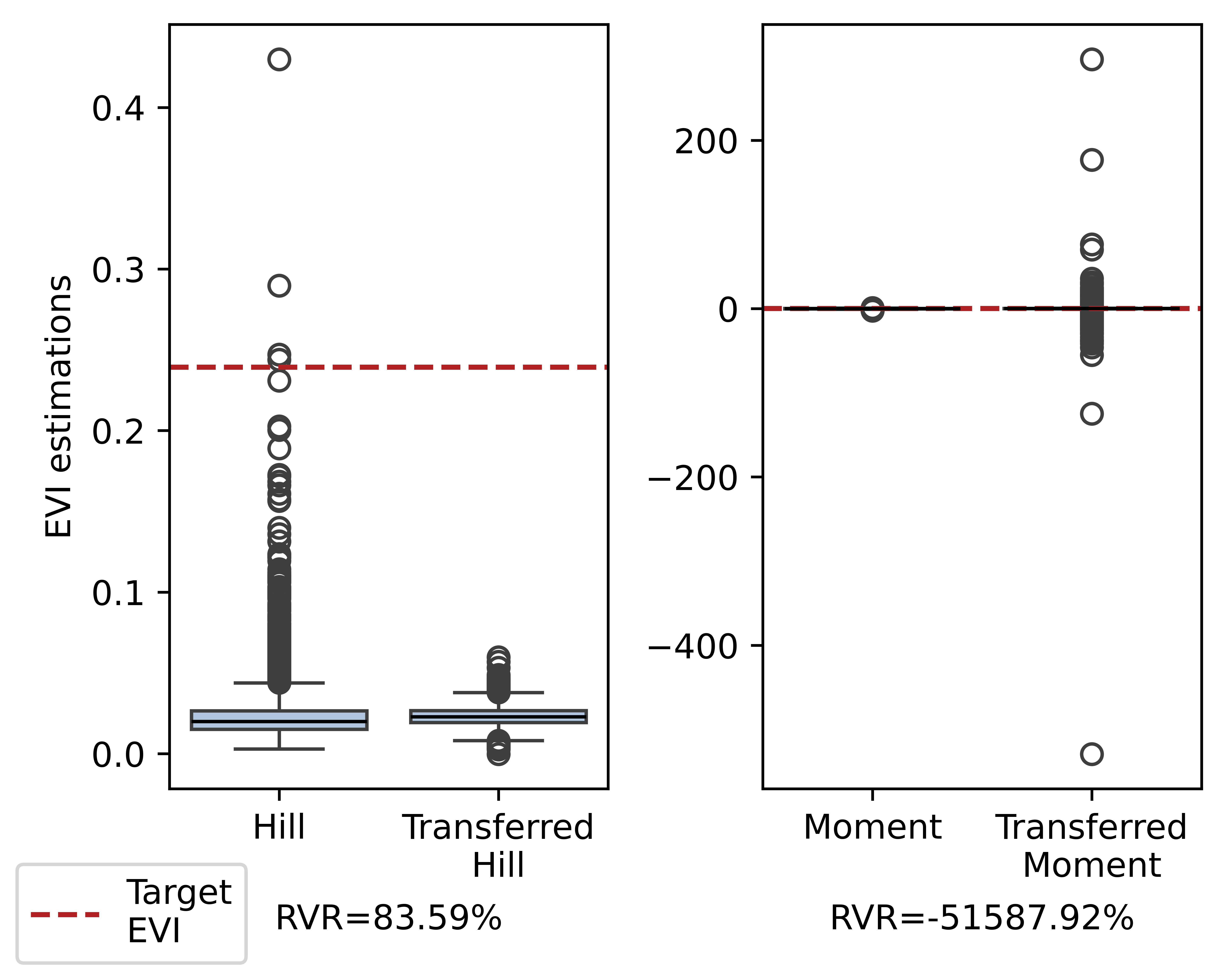}
        \captionsetup{justification=centering}
        \subcaption{$(n,k,m)=(100,10,5000)$ \\
        \textit{($\widehat{\corr}(A,B)=0.90$, $\widehat{\corr}(C,D)=0.75$, $\widehat{\lambda}=0.78$)}}
    \end{subfigure}
    \hfill
    \begin{subfigure}[b]{0.49\linewidth}
        \centering
        \includegraphics[height=0.25\textheight]{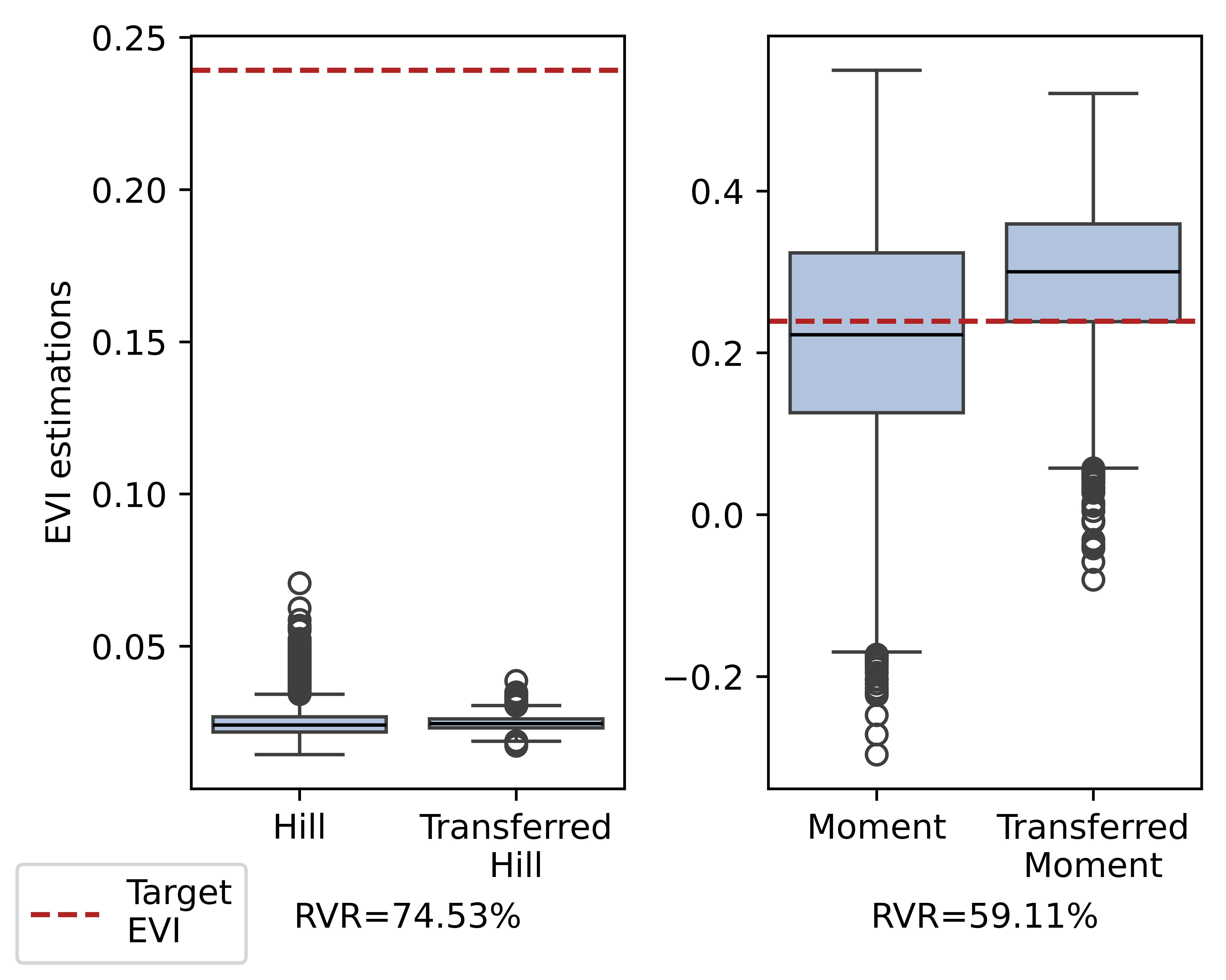}
        \captionsetup{justification=centering}
        \subcaption{$(n,k,m)=(1000,100,5000)$ \\
        \textit{($\widehat{\corr}(A,B)=0.95$, $\widehat{\corr}(C,D)=0.77$, $\widehat{\lambda}=0.79$)}}
    \end{subfigure}
    \caption{Boxplots of EVI estimations on the water surge dataset}
    \label{fig:flood_res}
\end{figure}

Two configurations are considered here: $n=100$ or $n=1,000$ coupled HF-BF samples, with $m=5,000$ additional BF samples. The results are presented in Figure \ref{fig:flood_res}.

The correlations and tail dependence estimations are similar in both cases, though the dependence seems to be lighter in the case $n=100$. This is likely due to the smaller sample size, which may exclude the most strongly dependent HF-LF observations. The transferred Hill estimator yields high RVR values, consistent with the results obtained on simulated data when the dependence between target and source is high. The variance reduction is stronger when $n=100$ as the baseline estimator has a larger variance to begin with. The bias, inherent to the baseline Hill estimator, is not increased.

The transferred moment estimator shows different results: when $n=1,000$, the variance is reduced but a bias appears, and when $n=100$ many outliers appear and the variance is increased. This can be explained by the choice of coefficients $(\alpha, \beta, \alpha', \beta')$, derived to reduce the variance of each moment and not that of the whole estimator, and their complex expressions composed of many terms to estimate. As said previously, the goal here is only to show that the transfer method can be applied to other EVI estimators; the coefficients of the Moment estimator would need to be optimized for the whole estimator rather than for each moment independently in order to guarantee variance reduction.

\subsection{Ice accretion model}
The ice accretion samples come from a physics-based simulation model \citep{trontin_description_2017}. Aircraft icing occurs when supercooled water droplets freeze upon contact with the surface of an aircraft, which modifies the geometry of the aircraft and can lead to performance degradation. It is mainly influenced by the angle of attack (the angle between the wing and the airflow), the airspeed, the ambient temperature, and the liquid water content. The ice accretion phenomenon has been involved in numerous fatal accidents and remains a critical area of aeronautics research.

\begin{figure}[!ht]
    \centering
    \includegraphics[width=0.4\linewidth]{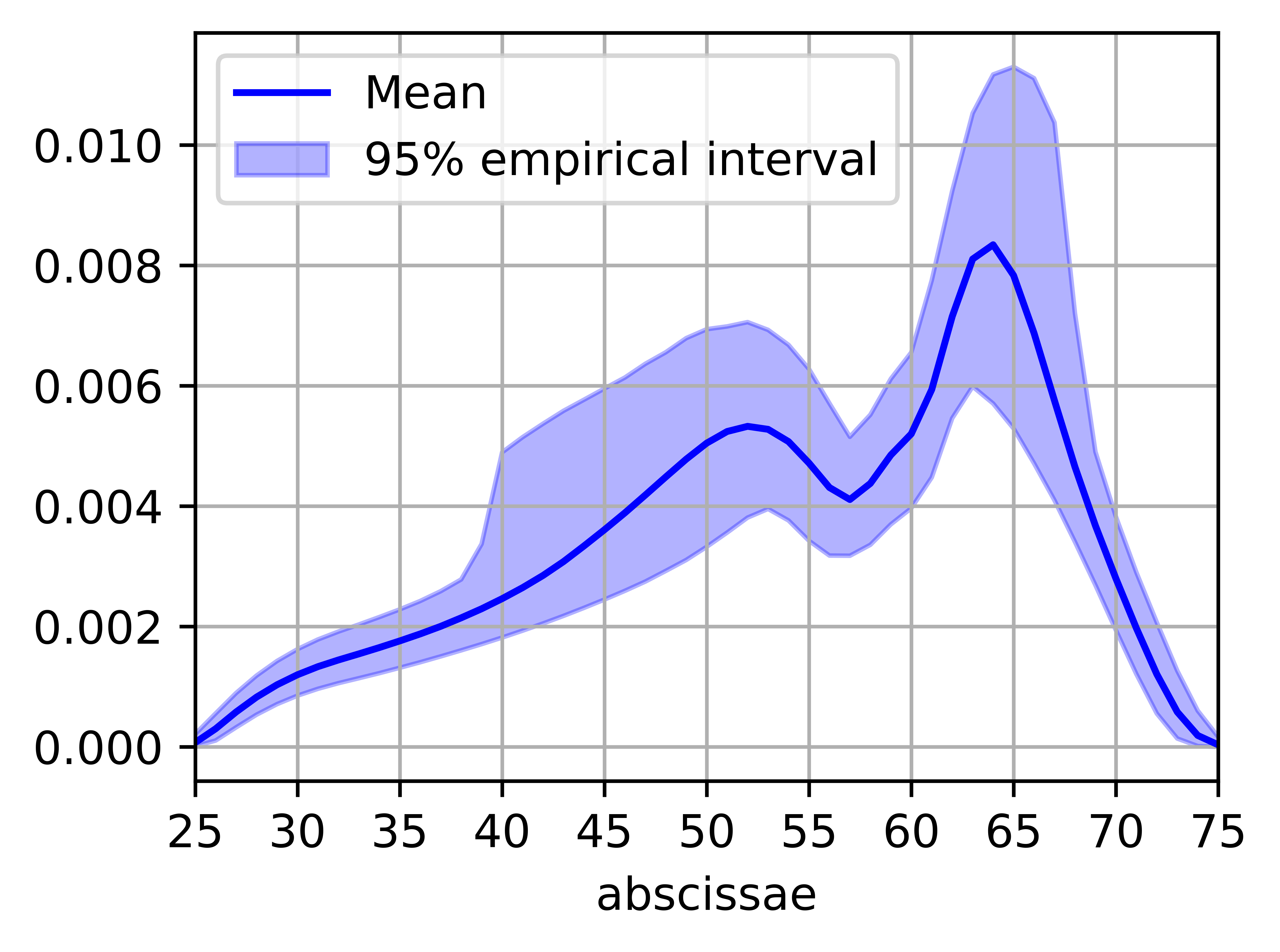}
    \caption{Mean and 95\% empirical interval for all profiles of ice accretion}
    \label{fig:ice_mean_interval}
\end{figure}

The dataset consists of 2,500 samples, each represented by a vector of 128 values measured along a curvilinear abscissa of the wing profile. In the following analysis, the source and target samples correspond to data at two different positions on the wing profile. Figure \ref{fig:ice_mean_interval} shows the mean ice accretion and its 95\% empirical interval at each abscissa, providing an overview of the data. The values for abscissae between 0 and 25, as well as between 75 and 128, are all zero and therefore excluded from the analysis. The parameters are set to $n=500$ coupled samples, $k=50$ extremes, and $m=2,000$ additional source samples. \\

\begin{figure}[!ht]
    \centering
    \includegraphics[width=0.6\linewidth]{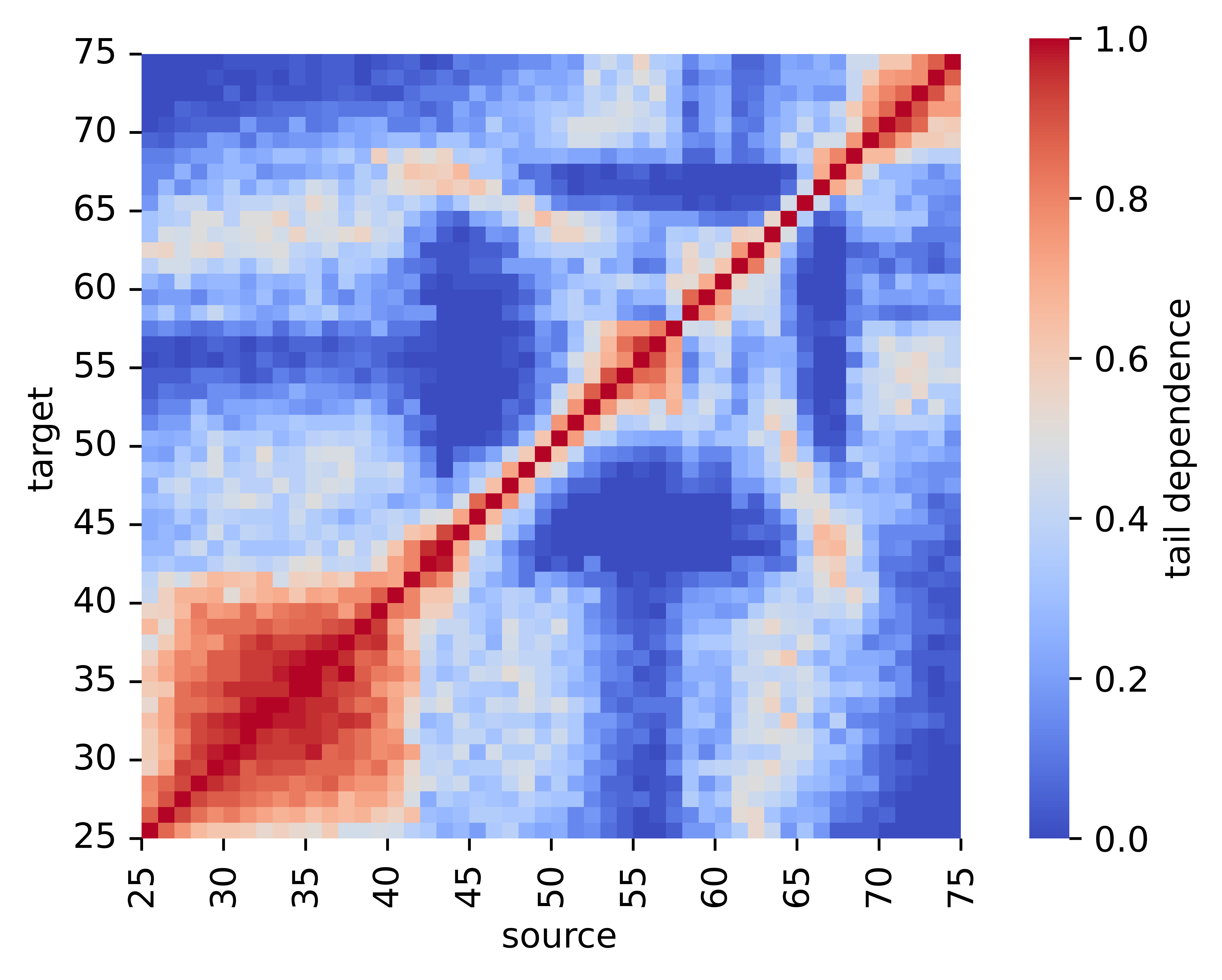}
    \caption{Tail dependence $\widehat{\lambda}$ for all combinations of abscissae}
    \label{fig:corr_ice}
\end{figure}

To choose which samples to consider as target and source, a grid search across all possible abscissae pairs is performed by computing the associated correlations $\widehat{\corr}(A,B)$ and $\widehat{\corr}(C,D)$ and the tail dependence $\widehat{\lambda}$. Pairs with high correlation and tail dependence can lead to strong variance reduction, allowing to illustrate the proposed method. The results of the grid search are presented in Figure \ref{fig:corr_ice}. The correlation heatmaps are not exactly symmetrical, as the source mean estimation is based on $n+m$ points, while the target mean estimation is based on only $n$ points. As expected, source and target samples exhibit strong dependence when they are spatially close on the profile, which can be seen on the main diagonal. Particularly high dependence is observed for abscissae 25 to 35, around 55, and from 70 to 75. Additionally, there are regions of moderate dependence even when the source and target samples are relatively distant, suggesting the presence of broader structural dependencies beyond local proximity. For the rest of the analysis, the following pairs of abscissae are considered: a target index at 36 and a source index at 31, representing closely located positions with strong correlation; and another with a target index at 67 and a source index at 43, representing more distant positions with moderate correlation. \\

\begin{figure}[!ht]
    \centering
    \begin{subfigure}[b]{\linewidth}
        \centering
        \includegraphics[width=0.8\linewidth]{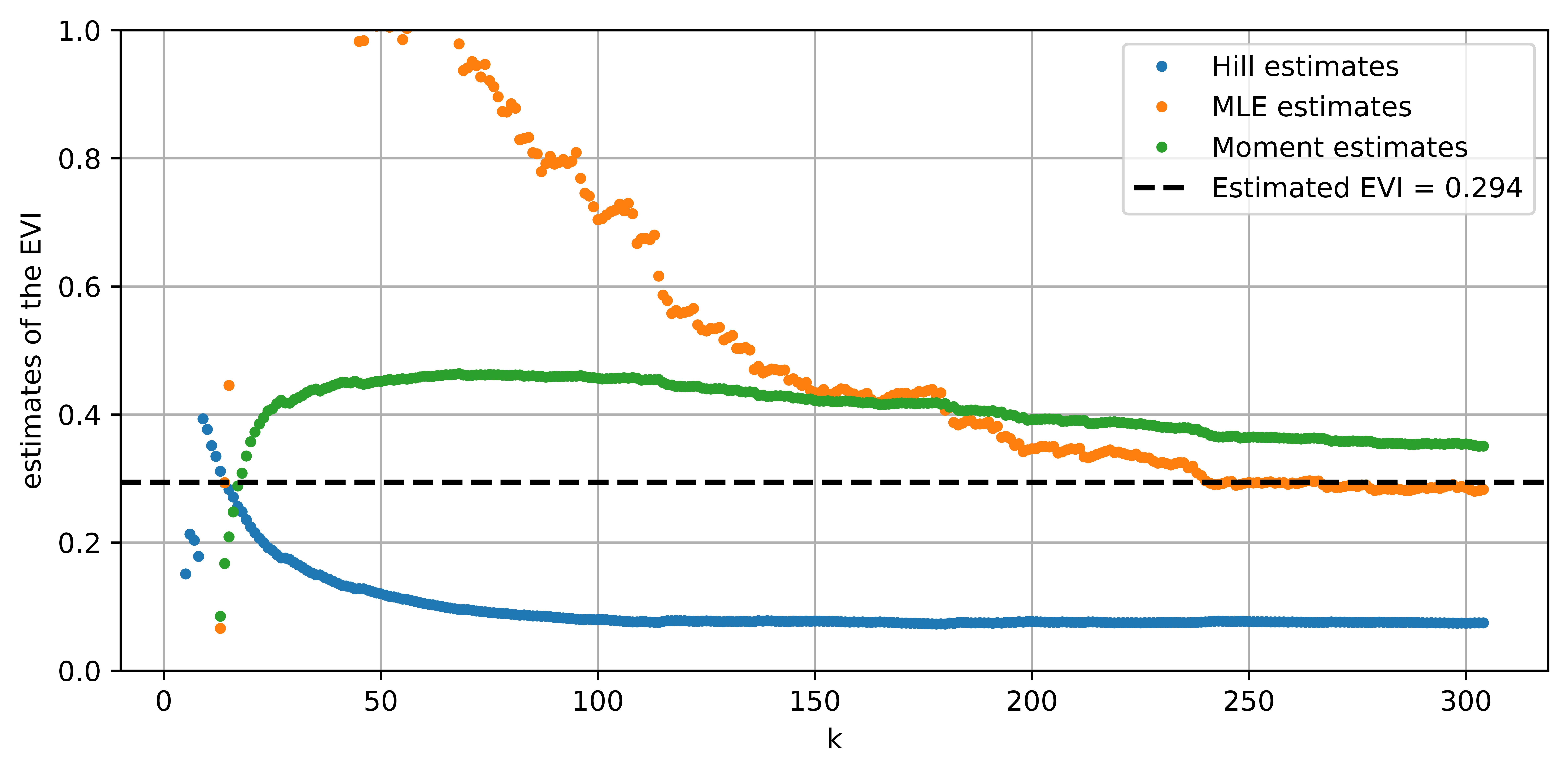}
        \subcaption{Target index=36}
    \end{subfigure}
    \hfill
    \begin{subfigure}[b]{\linewidth}
        \centering
        \includegraphics[width=0.8\linewidth]{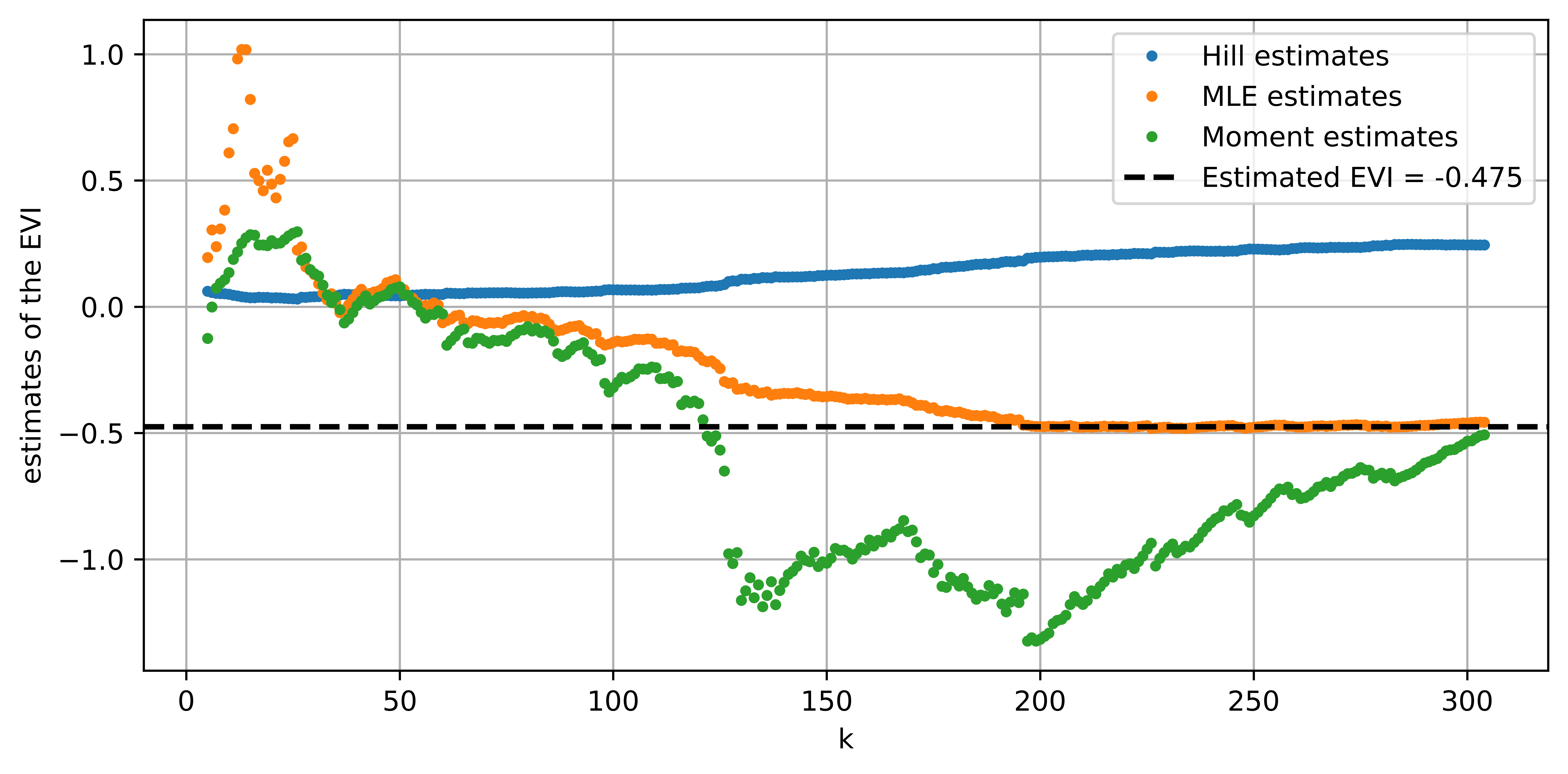}
        \subcaption{Target index=67}
    \end{subfigure}
    \caption{Hill plots on 2,500 ice accretion HF samples}
    \label{fig:hill_plots_ice}
\end{figure}

\begin{figure}[!ht]
    \centering
    \begin{subfigure}[b]{0.49\linewidth}
        \centering
        \includegraphics[height=0.25\textheight]{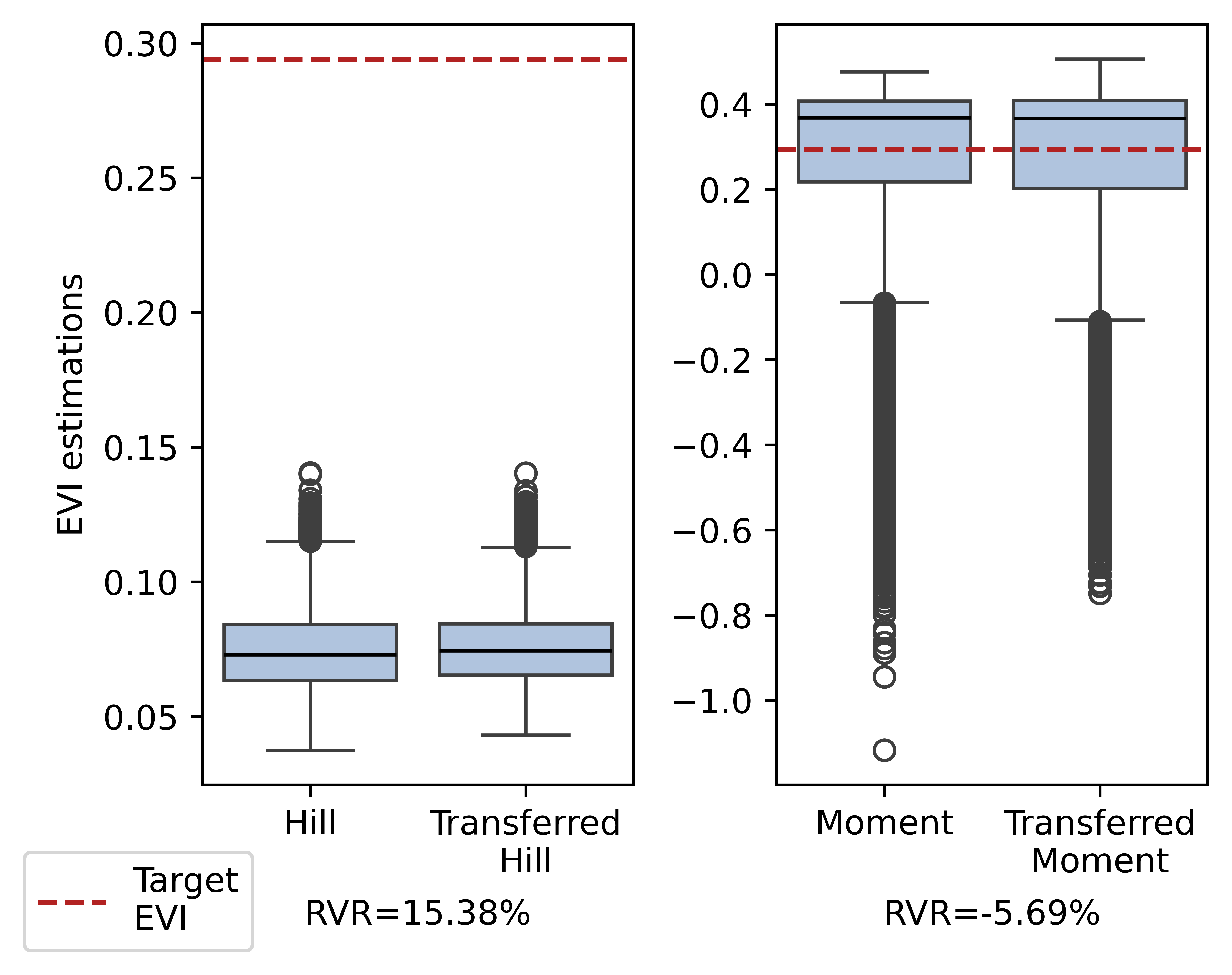}
        \captionsetup{justification=centering}
        \subcaption{Target index=36 and source index=31 \\
        \textit{($\widehat{\corr}(A,B)=0.72$, $\widehat{\corr}(C,D)=0.91$, $\widehat{\lambda}=0.92$)}}
    \end{subfigure}
    \hfill
    \begin{subfigure}[b]{0.49\linewidth}
        \centering
        \includegraphics[height=0.25\textheight]{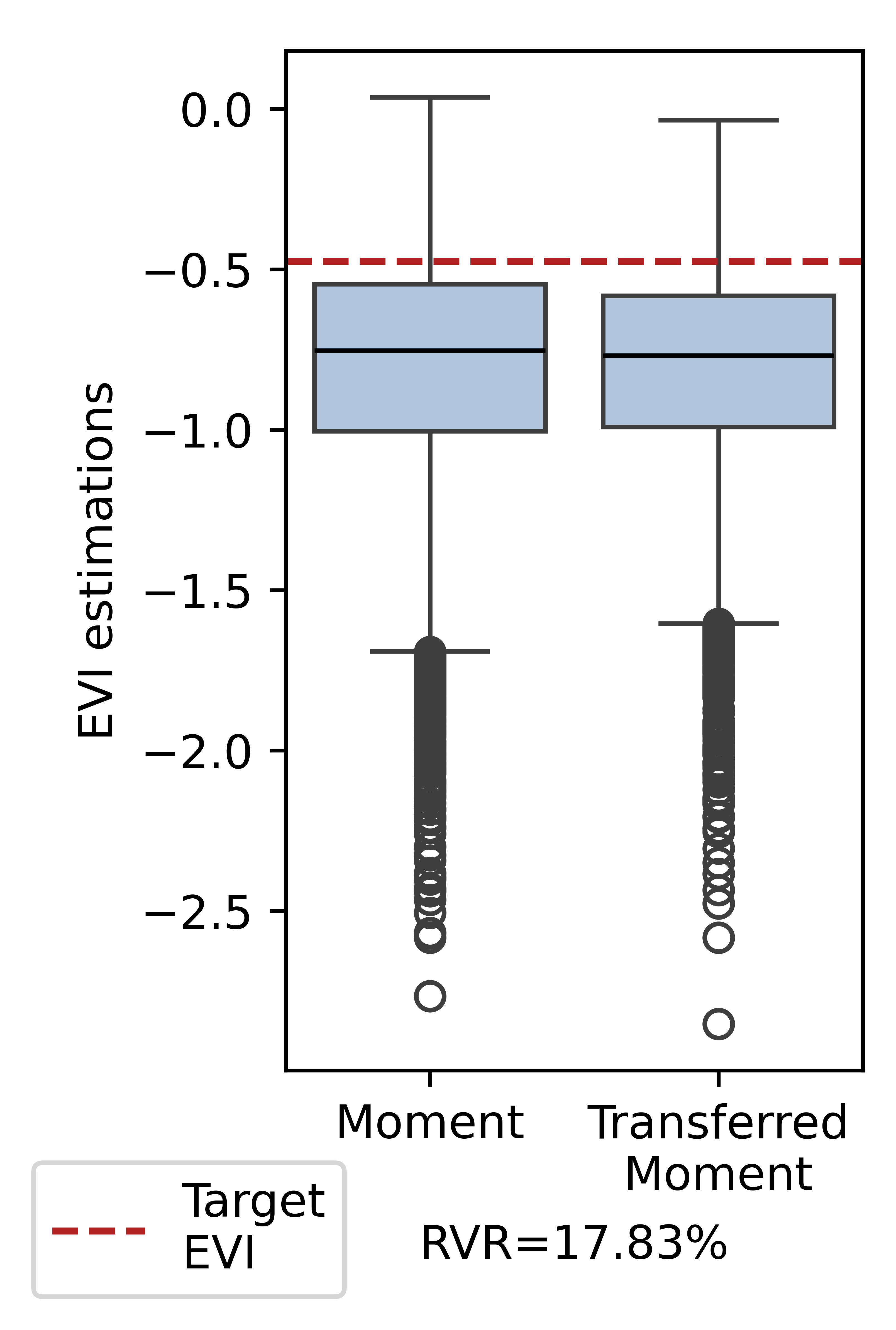}
        \captionsetup{justification=centering}
        \subcaption{Target index=67 and source index=43 \\
        \textit{($\widehat{\corr}(A,B)=0.65$, $\widehat{\corr}(C,D)=0.62$, $\widehat{\lambda}=0.66$)}}
    \end{subfigure}
    \caption{Boxplots of EVI estimations on the ice accretion dataset}
    \label{fig:boxplots_ice}
\end{figure}

The reference EVI values are computed using all 2,500 available target samples with the Hill plot  method illustrated in Figure \ref{fig:hill_plots_ice}. The target EVI for index 36 is estimated to be 0.294, and for index 67, -0.475. As noted in section \ref{sec:EVI_est}, the Hill estimator is known to suffer from bias, and this is observed in the estimates for both target indices 36 and 67 in the ice accretion dataset.

Figure \ref{fig:boxplots_ice} presents the results of the various EVI estimators for the two combinations of abscissae considered. The variance reduction obtained with the transferred Hill estimator is more moderate here, which can be explained by the limited number of additional source samples $m=2,000$. With the target index 67, the EVI to estimate is negative, and the transferred moment estimator offers a substantial variance reduction of 17\%.

\clearpage
\section{Conclusion}
In this paper, new variance-reduced EVI estimators using transfer learning have been introduced by applying the approximate control variates method for ratio of means estimator. The proposed transferred Hill estimator achieves significant variance reduction under strong tail dependence between the target and source distributions, as demonstrated theoretically and in simulations. Notably, under suitable assumptions, its asymptotic RVR is independent of the target and source EVIs, allowing the method to remain effective even when the two distributions differ substantially in tail heaviness. The variance reduction benefits from a larger number $m$ of additional source samples. Thanks to the fundamental property of control variates, the transferred Hill estimator necessarily yields a reduced, or at least non-increased, variance compared to the standard Hill estimator.

Two different applications were presented, on water surge multi-fidelity data and on ice accretion data, showing how widely the new estimators presented can be used. \\

A natural next step is to leverage the proposed variance-reduced EVI estimators for the estimation of extreme quantiles and rare event probabilities, which are critical in many risk assessment and engineering applications.

The proposed approach has been applied to the Hill estimator to estimate $\gamma^T>0$. We showed that it could be applied to the moment estimator to estimate $\gamma^T>-\frac{1}{2}$. However, its coefficients $\left(\alpha, \beta, \alpha', \beta'\right)$ should be optimized for the whole estimator rather than to minimize the variance of each moment independently, in order to get the variance reduction guarantee. Future work could aim to apply the approach to other EVI estimators, in particular for $\gamma^T\leq -\frac{1}{2}$.

\section*{Acknowledgements}
We gratefully acknowledge our ONERA colleagues Lokman Bennani, Ghislain Blanchard, Maxime Bouyges, and Patricia Klotz for providing the ice accretion dataset.

\section*{Declarations}
\subsection*{Supplementary material}
All codes used to produce the figures on simulated data are available on GitHub at \url{https://github.com/LouB-N/Variance-reduced-extreme-value-index-estimators.git}. 

\subsection*{Competing Interests}
 All authors certify that they have no affiliations with or involvement in any organization or entity with any financial interest or non-financial interest in the subject matter or materials discussed in this manuscript.

\subsection*{Declaration of generative AI use in the manuscript preparation process}
During the preparation of this work, the authors used Chat-GPT 5 in order to improve language and readability. After using this tool, the authors reviewed and edited the content as needed and take full responsibility for the content of the published article.
\vspace{20pt}

\appendix

\section{Proof of Proposition \ref{prop_consistent}: studying the consistence of the transferred Hill estimator} \label{sec:proof_consistent}
Since the samples of $B$ and $D$ are i.i.d. and their means are finite, the differences of Monte Carlo estimators on $n+m$ and on $n$ points almost surely tend to 0: \begin{align*}
\overline{B_{n+m}} - \overline{B_n} \overset{\text{a.s.}}{\underset{\substack{n \to \infty \\ m \to \infty}}{\longrightarrow}}
 0 \quad \text{ and } \quad \overline{D_{n+m}} - \overline{D_n} \overset{\text{a.s.}}{\underset{\substack{n \to \infty \\ m \to \infty}}{\longrightarrow}}
 0 \end{align*}

Therefore, the transferred Hill estimator is asymptotically equivalent to the Hill estimator: 
\begin{align*}
    \widehat{\gamma}^T_{TH} - \widehat{\gamma}^T_{T} = \frac{\overline{A_n} + \alpha (\overline{B_{n+m}} - \overline{B_n})}{\overline{C_n} + \alpha (\overline{D_{n+m}} - \overline{D_n})} - \frac{\overline{A_n}}{\overline{C_n}} \overset{\text{a.s.}}{\underset{\substack{n \to \infty \\ m \to \infty}}{\longrightarrow}} 0 \quad 
    \Longrightarrow \quad \widehat{\gamma}^T_{TH} - \widehat{\gamma}^T_{T} \overset{\mathbb{P}}{\underset{\substack{n \to \infty \\ m \to \infty}}{\longrightarrow}} 0
\end{align*}
Hence the transferred Hill estimator is asymptotically consistent under the same conditions as the Hill estimator, i.e. first and second order conditions for $m \to \infty$, $k = k(n) \to \infty$, $k/n \to 0$ as $n \to \infty$.

\section{Proof of Proposition \ref{prop}: studying the asymptotic variance reduction with the transferred Hill estimator} \label{sec:asymp_rvr}
In this section, the variance reduction with the transferred Hill estimator is studied. The following formula of variance reduction with the approximate control variates estimator compared to the baseline Monte Carlo estimator for a ratio of means comes from \citep{bocquet_control_2025}.
\begin{align*} \var\left(\widehat{\gamma}^T_{H}\right) - \var\left(\widehat{\gamma}^T_{TH}\right) \underset{n \to \infty}{\approx} - \frac{m}{n(n+m)}\frac{1}{\E[C]^2}
\frac{Var\bigg(\big(\gamma^T\cov(B,C)-\cov(A,B)\big)D - \big(\gamma^T\cov(C,D)-\cov(A,D)\big)B\bigg)}{\var(B)\var(D)-\cov(B,D)^2}
\end{align*}
\begin{align*}
\text{where } &A = (\ln(Y^T) - \ln(Y^T_{n-k:n})) \1_{\{Y^T > Y^T_{n-k:n}\}} \; &&B = (\ln(Y^S) - \ln(Y^S_{n-k:n})) \1_{\{Y^S > Y^S_{n-k:n}\}} \\
&C = \1_{\{Y^T > Y^T_{n-k:n}\}} \; &&D = \1_{\{Y^S > Y^S_{n-k:n}\}}\end{align*}

We want to study the variance reduction asymptotically, that is when $k=k(n) \to \infty$, $\frac{k}{n} \to 0$ for $n\to \infty$. That means the thresholds $Y^T_{n-k:n}$ and $Y^S_{n-k:n}$ tend to the target and source endpoints $y^{T*}$ and $y^{S*}$. For the sake of simplicity, we note the limits as $n\to \infty$. \\

\subsection{Preliminary results}
As stated in Theorem~\ref{th:POT} (Pickands–Balkema–de Haan), the conditional excess distribution of a random variable $Y$ over a threshold $u$ converges to the Generalized Pareto distribution:
\begin{align*}F \in \mathcal{D}(GEV_{\gamma}) \Longleftrightarrow& \underset{u \to y^*}{\text{lim}} F_u\left(\sigma(u) y\right) = GP_{\gamma}(y)\\ &:= \begin{cases} 1 - (1+\gamma y)^{-\frac{1}{\gamma}} \quad &\text{if } \gamma \neq 0, \quad 0<y<(0\vee(-\gamma))^{-1}; \\ 1-exp(-y) \quad &\text{if } \gamma=0,\quad  y \in \mathbb{R} \end{cases}\end{align*}
where $F_u(y)=\mathbb{P}(Y-u\leq y|Y>u)$ and $\sigma(\cdot)$ a positive scaling function. \\

To analyze the asymptotic variance reduction, it is necessary to study the asymptotic distribution of the log-excesses. Observe the relationship between the distribution of excesses and that of the log-excesses:
\begin{align*}
    \mathbb{P}\left(\ln\left(\frac{Y}{u}\right) \leq y \bigg|Y >u\right) &= \mathbb{P}\left(\frac{Y}{u} \leq e^y \bigg|Y >u\right)\\
    &= \mathbb{P}\left(Y - u \leq u(e^y-1) |Y >u\right) \\
    &= F_u\left( u(e^y-1) \right)\\
    &\underset{u \to y^*}{\approx} GP_{\gamma}\left( \frac{1}{\gamma}(e^y-1) \right) \quad \text{if } \gamma>0
\end{align*}

This allows to compute the asymptotic mean log-excess. The asymptotic cdf is $GP_{\gamma}\left( \frac{1}{\gamma}(e^y-1) \right), \forall y \geq 0$, therefore the asymptotic pdf is\begin{align*}
    \frac{d}{dy}GP_{\gamma}\left( \frac{1}{\gamma}(e^y-1) \right) = \frac{d}{dy} \left( 1 - e^{-\frac{y}{\gamma}} \right) = \frac{1}{\gamma}e^{-\frac{y}{\gamma}}
\end{align*}
We note that this is the pdf of an exponential distribution with parameter $1/\gamma$.\\
The asymptotic mean log-excess is derived as:
\begin{align*}
    \E\left[\ln\left(\frac{Y}{u}\right)\Bigg|Y >u\right] \underset{u \to y^*}{\approx} \E\left[X\right] &= \int_{0}^{+ \infty} \frac{y}{\gamma}e^{-\frac{y}{\gamma}} dy \quad \text{where } X \text{ has cdf } GP_{\gamma}\left( \frac{1}{\gamma}(e^y-1) \right)\\
    &= \gamma \int_{0}^{+ \infty} z e^{-z} dz \quad \text{with the change of variable }z=\frac{y}{\gamma}\\
    &= \gamma \\
\end{align*}

To conclude, the asymptotic mean log-excess is: \begin{align} \label{eq:log_excess}
    \E\left[\ln\left(\frac{Y}{u}\right)\Bigg|Y >u\right] \underset{u \to y^*}{\approx}
        \gamma \quad &\text{if } \gamma >0
\end{align}
Note that when $\gamma>0$, the log-excesses are asymptotically exponentially distributed with mean $\gamma$, which implies that the scaled log-excesses $Z := \left( \frac{\ln(Y/u)}{\gamma} \bigg| Y>u\right)$ asymptotically follow a standard exponential distribution.

\subsection{Studying the asymptotic variance reduction}
\subsubsection{Proof of the first statement in Proposition \ref{prop}}
Define $\epsilon_A = A - \gamma^T C$, the random deviation of the log excesses from it expected value, with the useful property $\E[\epsilon_A] = \E[A] - \gamma^T \E[C] \underset{n \to \infty}{\rightarrow} 0$, as $\frac{\E[A]}{\E[C]} = \E[\ln(Y^T) - \ln(Y^T_{n-k:n}) | Y^T > Y^T_{n-k:n}] \underset{n \to \infty}{\rightarrow} \gamma^T$ (see Remark 1.2.3 of \citep{de_haan_extreme_2006} or previous section).\\

By replacing $A$ with $\epsilon_A + \gamma^T C$,the following terms are simplified:
\begin{align*}
    \gamma^T\cov(B,C)-\cov(A,B) &= \gamma^T\cov(B,C)-\cov(\epsilon_A + \gamma^T C,B) = - \cov(\epsilon_A,B) \\
    \text{and } \gamma^T\cov(C,D)-\cov(A,D) &= \gamma^T\cov(C,D)-\cov(\epsilon_A + \gamma^T C,D) = - \cov(\epsilon_A,D)
\end{align*}

The variance expression can be written as:
\begin{align*}
    &\var\left(\widehat{\gamma}^T_{H}\right) - \var\left(\widehat{\gamma}^T_{TH}\right)\\
    \underset{n \to \infty}{\approx}& \frac{m}{n(n+m)}\frac{1}{\E[C]^2} \frac{Var\bigg(- \cov(\epsilon_A,B) D + \cov(\epsilon_A,D)B\bigg)}{\var(B)\var(D)-\cov(B,D)^2} \\
    \underset{n \to \infty}{\approx}& \frac{m}{n(n+m)}\frac{1}{\E[C]^2} \frac{\cov(\epsilon_A,B)^2 \var(D) + \cov(\epsilon_A,D)^2 \var(B) - \cov(\epsilon_A,B)\cov(\epsilon_A,D)\cov(B,D)}{\var(B)\var(D)-\cov(B,D)^2}
\end{align*} \\

Define $\epsilon_A' = \epsilon_A/\gamma^T$ that is asymptotically independent from $\gamma^T$. Indeed, $\epsilon_A' = (\frac{\ln(Y^T) - \ln(Y^T_{n-k:n})}{\gamma^T} - 1) \1_{\{Y^T > Y^T_{n-k:n}\}} = (Z - 1)/\1_{\{Y^T > Y^T_{n-k:n}\}}$ where $Z$ asymptotically follows a standard exponential distribution, therefore $\epsilon_A'$  is asymptotically independent from $\gamma^T$.

The variance reduction can be written as:
\begin{align*}
    &\var\left(\widehat{\gamma}^T_{H}\right) - \var\left(\widehat{\gamma}^T_{TH}\right) \\
    \underset{n \to \infty}{\approx}& (\gamma^T)^2 \frac{m}{n(n+m)}\frac{1}{\E[C]^2} \frac{\cov(\epsilon_A',B)^2 \var(D) + \cov(\epsilon_A',D)^2 \var(B) - \cov(\epsilon_A',B)\cov(\epsilon_A',D)\cov(B,D)}{\var(B)\var(D)-\cov(B,D)^2}\\
    :=& (\gamma^T)^2 W
\end{align*}

Considering the relative variance reduction:\begin{align*}
    RVR = \frac{\var\left(\widehat{\gamma}^T_{H}\right) - \var\left(\widehat{\gamma}^T_{TH}\right)}{\var\left(\widehat{\gamma}^T_{H}\right)} \underset{n \to \infty}{\approx}W \qquad \text{as } \var\left(\widehat{\gamma}^T_{H}\right) \underset{n \to \infty}{\approx} (\gamma^T)^2
\end{align*}
To prove the RVR is asymptotically independent from $\gamma^T$, we need to prove $W$ is asymptotically independent from $\gamma^T$. The variables $B$ and $D$ are independent of $\gamma^T$ as they are functions of only $Y^S$. We showed $\epsilon_A'$ is asymptotically independent from $\gamma^T$. The variable $C$ follows a Bernoulli distribution with success probability $p = k/n$, independent from $\gamma^T$. Considering all the random variables in $W$ are asymptotically independent from $\gamma^T$, so is $W$. \\

\subsubsection{Proof of the second statement in Proposition \ref{prop}}
Let us study $W$ in more detail. \textbf{Assume $Y^S \in \mathcal{D}(G_{\gamma^S})$ and $\gamma^S>0$ for now.} We will need the following expressions: \begin{align*}
&\bullet \E[C] = \E[D] := p \\
&\bullet \E[B] = \E[\ln(Y^S) - \ln(Y^S_{n-k:n}) | Y^S > Y^S_{n-k:n}] \mathbb{P}(Y^S > Y^S_{n-k:n}) \underset{n \to \infty}{\approx} \gamma^S p \\
&\bullet \E[B^2] = \E[\left(\ln(Y^S) - \ln(Y^S_{n-k:n})\right)^2 | Y^S > Y^S_{n-k:n}] \mathbb{P}(Y^S > Y^S_{n-k:n}) \underset{n \to \infty}{\approx} \E[Z^2] p = 2 (\gamma^S)^2 p \\ &\quad \text{where } Z \text{ is a standard exponential r.v.}\\
&\bullet \E[BD] = \E[B] \underset{n \to \infty}{\approx} \gamma^S p \\
&\bullet \E[\epsilon_A'] \underset{n \to \infty}{\approx} 0 \\
&\bullet \E[\epsilon_A'D] = E\left[ \frac{\ln(Y^T) - \ln(Y^T_{n-k:n})}{\gamma^T} - 1 \Bigg| Y^T > Y^T_{n-k:n}, Y^S > Y^S_{n-k:n}\right] \mathbb{P}(Y^T > Y^T_{n-k:n}, Y^S > Y^S_{n-k:n}) \\ &\hspace{35pt} \underset{n \to \infty}{\approx} c_{AD} \lambda p \\
&\bullet \E[\epsilon_A'B] = \E\left[ \left( \frac{\ln(Y^T) - \ln(Y^T_{n-k:n})}{\gamma^T} - 1 \right) \left(\ln(Y^S) - \ln(Y^S_{n-k:n})\right) \Bigg| Y^T > Y^T_{n-k:n}, Y^S > Y^S_{n-k:n}\right] \\ & \hspace{50pt} \times \mathbb{P}(Y^T > Y^T_{n-k:n}, Y^S > Y^S_{n-k:n}) \underset{n \to \infty}{\approx} \gamma^S c_{AB} \lambda p
\end{align*}
where $\lambda$ is the tail dependence. \begin{align*}
c_{AD} &= E\left[ \frac{\ln(Y^T) - \ln(Y^T_{n-k:n})}{\gamma^T} - 1 \Bigg| Y^T > Y^T_{n-k:n}, Y^S > Y^S_{n-k:n}\right] \\ &\underset{n \to \infty}{\approx} E\left[ Z - 1 \Bigg| Y^T > Y^T_{n-k:n}, Y^S > Y^S_{n-k:n}\right] \\
c_{AB} &= \E\left[ \left( \frac{\ln(Y^T) - \ln(Y^T_{n-k:n})}{\gamma^T} - 1 \right) \left(\frac{\ln(Y^S) - \ln(Y^S_{n-k:n})}{\gamma^S}\right) \Bigg| Y^T > Y^T_{n-k:n}, Y^S > Y^S_{n-k:n}\right] \\ &\underset{n \to \infty}{\approx} E\left[ (Z - 1)Z \Bigg| Y^T > Y^T_{n-k:n}, Y^S > Y^S_{n-k:n}\right]\end{align*} 

$c_{AD}$ and $c_{AB}$ are measures of extremal dependence. Considering $Z$ is a standard exponential variable, independent from $\gamma^T$ and $\gamma^S$, then $c_{AD}$ and $c_{AB}$ are asymptotically independent from $\gamma^T$ and $\gamma^S$.\\

To rewrite the terms in the variance reduction in an asymptotic form:
\begin{align*}
    &\bullet \var(D) = p(1-p) \\
    &\bullet \var(B) = \E[B^2] - \E[B]^2 \underset{n \to \infty}{\approx} 2 (\gamma^S)^2 p - (\gamma^S)^2 p^2 =  (\gamma^S)^2 p (2-p)\\
    &\bullet \cov(B,D) = \E[BD] - \E[B]\E[D] \underset{n \to \infty}{\approx} \gamma^S p - \gamma^S p^2 = \gamma^S p (1-p) \\
    &\bullet \cov(\epsilon_A',D) = \E[\epsilon_A'D] - \E[\epsilon_A']\E[D] \underset{n \to \infty}{\approx} c_{AD} \lambda p \\
    &\bullet \cov(\epsilon_A',B) = \E[\epsilon_A'B] - \E[\epsilon_A']\E[B] \underset{n \to \infty}{\approx} \gamma^S c_{AB} \lambda p 
\end{align*}

Using these expressions, we show the RVR is asymptotically independent from $\gamma^T$ and $\gamma^S>0$:
\begin{align*}
    RVR &\underset{n \to \infty}{\approx} \frac{m}{n(n+m)} \frac{1}{p^2} \frac{\cov(\epsilon_A',B)^2 \var(D) + \cov(\epsilon_A',D)^2 \var(B) - \cov(\epsilon_A',B)\cov(\epsilon_A',D)\cov(B,D)}{\var(B)\var(D)-\cov(B,D)^2}\\
    &\underset{n \to \infty}{\approx} \frac{m}{n(n+m)} \frac{(\gamma^S)^2 c_{AB}^2 \lambda^2 p^3 (1-p) + c_{AD}^2 \lambda^2 p^3 (2-p) (\gamma^S)^2 - c_{AB} c_{AD} \lambda^2 p^3 (1-p) (\gamma^S)^2}{p^2 \left( (\gamma^S)^2 p^2 (2-p) (1-p) - (\gamma^S)^2p^2 (1-p)^2 \right)} \\
    &\underset{n \to \infty}{\approx} \lambda^2 \frac{m}{n(n+m)} \frac{c_{AB}^2 + c_{AD}^2 \frac{2-p}{1-p} - c_{AB} c_{AD}}{p}
\end{align*} \\
We can also clearly see a link with the tail dependence, even though the RVR also depends on $c_{AB}$ and $c_{AD}$, which also measure a form of extremal dependence. \\

\vspace{10pt}
\section{Selecting the source threshold} \label{sec:seuil}
\begin{figure}[!ht]
    \centering
    \begin{subfigure}[b]{0.49\linewidth}
        \centering
        \includegraphics[width=\linewidth]{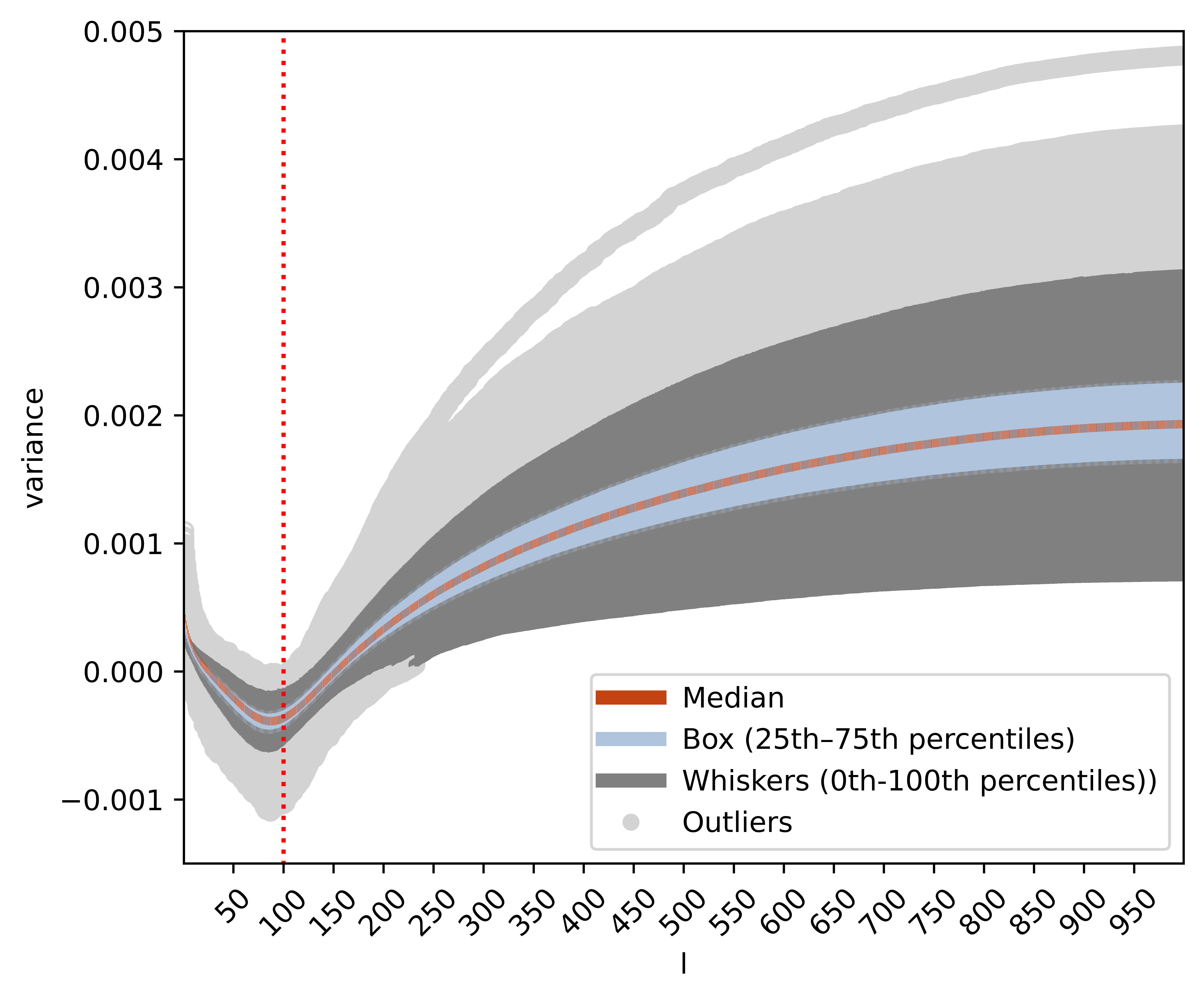}
    \end{subfigure}
    \hfill
    \begin{subfigure}[b]{0.49\linewidth}
        \centering
        \includegraphics[width=\linewidth]{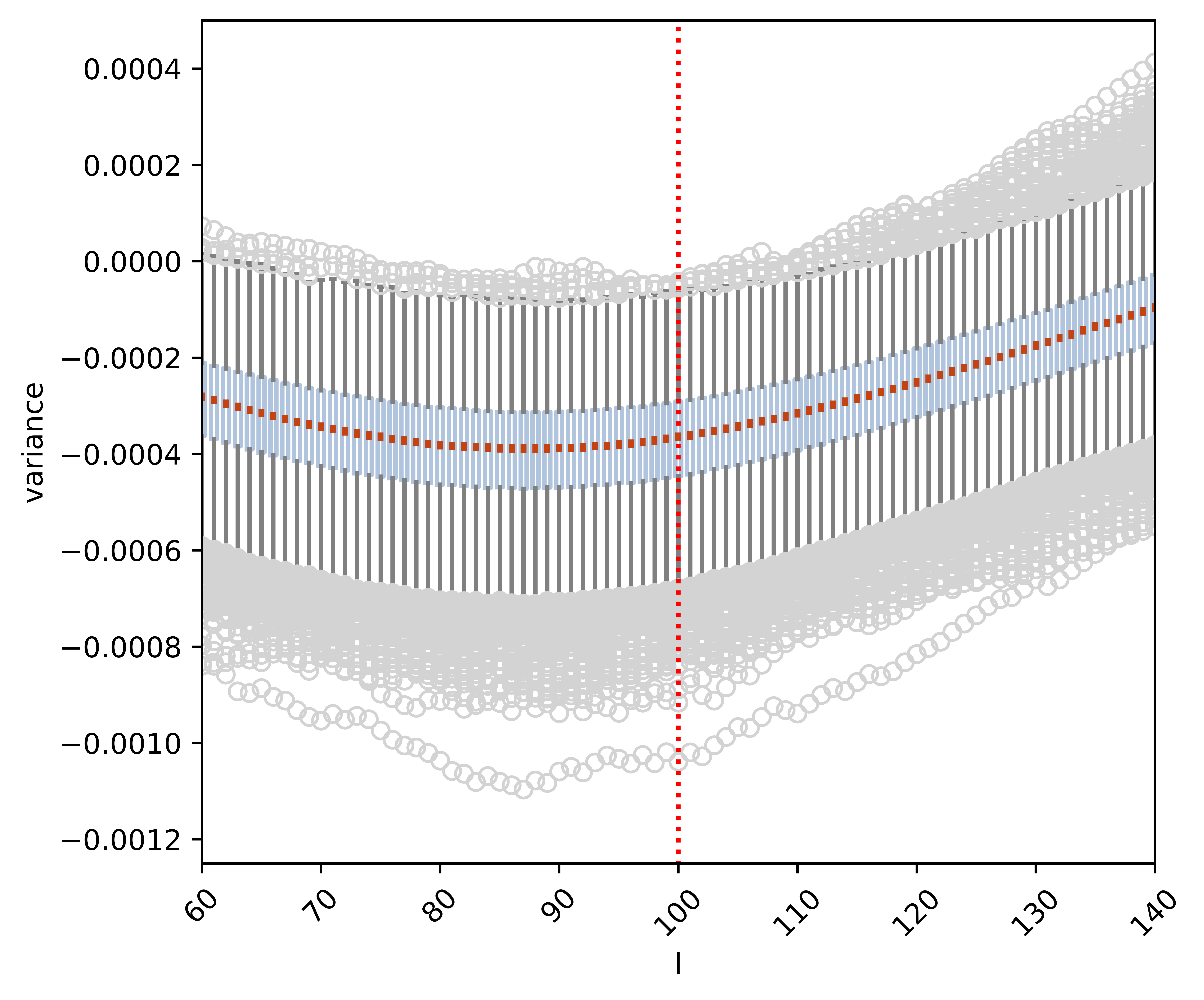}
    \end{subfigure}
    \captionsetup{justification=centering}
    \caption{Boxplots of the analytical variance of the transferred Hill estimator for different numbers of source extremes $l$\\
    \footnotesize{Left: full range of $l \in \{1, 999\}$. Right: zoomed-in view for $l \in \{60, 140\}$. The red dotted line marks $k = 100$.}}
    \label{fig:var_threshold}
\end{figure}

The goal here is to determine the optimal number of source extremes $l$, that is the optimal source threshold, given a fixed number of target extremes $k=0.1n$, using simulated data. As in Section \ref{sec:simu}, the data is generated using Pareto marginals with $\gamma^T=0.25$ and $\gamma^S=0.5$, a Gumbel copula with parameter $\theta=5$, $n=1,000$ coupled target-source samples, $k=100$ target extremes, and $m=5,000$ additional source data. The results of 10,000 simulations are presented in Figure \ref{fig:var_threshold}.

The analytical variance of the transferred Hill estimator is computed in each of the 10,000 simulations, and the resulting values are displayed in Figure \ref{fig:var_threshold}. Some variance estimates appear negative, particularly for small values of $l$, which may be due to the limited number of extreme observations used in the estimation: it makes the analytical expression unstable and unreliable.

The minimum estimated variance is found slightly below $l = 100$, which coincides with the number of target extremes $k=100$. However, as seen in the right plot, differences in variance values are relatively small in this region.

When the dependence between the source and target is weaker, the optimal number of source extremes minimizing variance tends to shift slightly upward. Nevertheless, using the same number of extremes for both source and target samples remains a reasonable and practical choice. Setting $l=k$ simplifies the analysis and retains nearly optimal performance across different dependence structures.

\bibliographystyle{apalike}
\bibliography{sn-bibliography}

@article{caeiro_threshold_2015,
	title = {Threshold selection in extreme value analysis},
	journal = {Extreme value modeling and risk analysis: Methods and applications},
	author = {Caeiro, Frederico and Gomes, M. Ivette},
	year = {2015},
	pages = {69--87},
    doi={https://doi.org/10.1201/b19721-8},
}

@article{ahmed_extreme_2025,
	title = {Extreme {Value} {Statistics} in {Semi}-{Supervised} {Models}},
	volume = {120},
	issn = {0162-1459, 1537-274X},
	doi = {https://doi.org/10.1080/01621459.2024.2333582},
	number = {549},
	journal = {Journal of the American Statistical Association},
	author = {Ahmed, Hanan and Einmahl, John H.J. and Zhou, Chen},
	year = {2025},
	pages = {291--304},
}

@misc{kim_parametric_2024,
	title = {Parametric multi-fidelity {Monte} {Carlo} estimation with applications to extremes},
	doi = {https://doi.org/10.48550/arXiv.2410.08523},
	publisher = {arXiv},
	author = {Kim, Minji and Brown, Brendan and Pipiras, Vladas},
	year = {2024},
}

@book{de_haan_extreme_2006,
	series = {Springer {Series} in {Operations} {Research} and {Financial} {Engineering}},
	title = {Extreme {Value} {Theory}},
	isbn = {978-0-387-23946-0 978-0-387-34471-3},
	publisher = {Springer},
	author = {de Haan, Laurens and Ferreira, Ana},
	year = {2006},
	doi = {https://doi.org/10.1007/0-387-34471-3},
}

@inproceedings{fisher_limiting_1928,
	title = {Limiting forms of the frequency distribution of the largest or smallest member of a sample},
	volume = {24},
	booktitle = {Mathematical proceedings of the {Cambridge} philosophical society},
	publisher = {Cambridge University Press},
	author = {Fisher, Ronald Aylmer and Tippett, Leonard Henry Caleb},
	year = {1928},
	pages = {180--190},
    doi={https://doi.org/10.1017/S0305004100015681}
}

@article{gnedenko_sur_1943,
	title = {Sur la distribution limite du terme maximum d'une s\'erie aleatoire},
	journal = {The {Annals} of {Mathematics}},
	author = {Gnedenko, Boris},
	year = {1943},
	pages = {423--453},
    doi={https://doi.org/10.2307/1968974}
}

@article{balkema_residual_1974,
	title = {Residual life time at great age},
	journal = {The {Annals} of {Probability}},
	author = {Balkema, August A. and de Haan, Laurens},
	year = {1974},
	pages = {792--804},
    doi={https://doi.org/10.1214/aop/1176996548}
}

@article{pickands_iii_statistical_1975,
	title = {Statistical inference using extreme order statistics},
	journal = {The {Annals} of {Statistics}},
	author = {{Pickands III}, James},
	year = {1975},
	pages = {119--131},
    doi={https://doi.org/10.1214/aos/1176343003}
}

@article{hill_simple_1975,
	title = {A simple general approach to inference about the tail of a distribution},
	journal = {The {Annals} of {Statistics}},
	author = {Hill, Bruce M.},
	year = {1975},
	pages = {1163--1174},
    doi={https://doi.org/10.1214/aos/1176343247}
}

@incollection{de_haan_slow_1984,
	title = {Slow {Variation} and {Characterization} of {Domains} of {Attraction}},
	isbn = {978-90-481-8401-9 978-94-017-3069-3},
	booktitle = {Statistical {Extremes} and {Applications}},
	publisher = {Springer Netherlands},
	author = {de Haan, Laurens},
	year = {1984},
	doi = {https://doi.org/10.1007/978-94-017-3069-3\_4},
}

@article{dekkers_moment_1989,
	title = {A moment estimator for the index of an extreme-value distribution},
	journal = {The {Annals} of {Statistics}},
	author = {Dekkers, Arnold LM and Einmahl, John H.J. and de Haan, Laurens},
	year = {1989},
	note = {Publisher: JSTOR},
	pages = {1833--1855},
    doi={https://doi.org/10.1214/aos/1176347397}
}

@book{bousquet_extreme_2021,
	title = {Extreme {Value} {Theory} with {Applications} to {Natural} {Hazards}: {From} {Statistical} {Theory} to {Industrial} {Practice}},
	isbn = {978-3-030-74941-5 978-3-030-74942-2},
	publisher = {Springer International Publishing},
	editor = {Bousquet, Nicolas and Bernardara, Pietro},
	year = {2021},
	doi = {https://doi.org/10.1007/978-3-030-74942-2},
}

@book{embrechts_modelling_1997,
	title = {Modelling {Extremal} {Events}},
	isbn = {978-3-642-08242-9 978-3-642-33483-2},
	publisher = {Springer Berlin Heidelberg},
	author = {Embrechts, Paul and Klüppelberg, Claudia and Mikosch, Thomas},
	year = {1997},
	doi = {https://doi.org/10.1007/978-3-642-33483-2},
}

@article{zhan_predicting_2015,
	title = {Predicting cyber attack rates with extreme values},
	journal = {IEEE Transactions on Information Forensics and Security},
	author = {Zhan, Zhenxin and Xu, Maochao and Xu, Shouhuai},
	year = {2015},
	pages = {1666--1677},
    doi={http://doi.org/10.1109/TIFS.2015.2422261}
}

@article{einmahl_limits_2019,
	title = {Limits to {Human} {Life} {Span} {Through} {Extreme} {Value} {Theory}},
	issn = {0162-1459, 1537-274X},
	doi = {http://doi.org/10.1080/01621459.2018.1537912},
	journal = {Journal of the American Statistical Association},
	author = {Einmahl, Jesson J. and Einmahl, John H. J. and de Haan, Laurens},
	year = {2019},
	pages = {1075--1080},
}

@article{delaluz_estimating_2018,
	title = {Estimating the {Maximum} {Intensities} of {Soft} {X}-{Ray} {Flares} {Using} {Extreme} {Value} {Theory}},
	issn = {0038-0938, 1573-093X},
	doi = {http://doi.org/10.1007/s11207-018-1342-1},
	journal = {Solar Physics},
	author = {{De la Luz}, V. and Balanzario, E. P. and Tsiftsi, T.},
	year = {2018},
}

@article{asmussen_extreme_1998,
	title = {Extreme {Value} {Theory} for {Queues} {Via} {Cycle} {Maxima}},
	doi = {https://doi.org/10.1023/A:1009970005784},
	journal = {Extremes},
	author = {Asmussen, Søren},
	year = {1998},
	pages = {137--168},
}

@incollection{asmussen_variance-reduction_2007,
	title = {Variance-{Reduction} {Methods}},
	isbn = {978-0-387-30679-7 978-0-387-69033-9},
	booktitle = {Stochastic {Modelling} and {Applied} {Probability}},
	publisher = {Springer New York},
	author = {Asmussen, Søren and Glynn, Peter W.},
	year = {2007},
	doi = {https://doi.org/10.1007/978-0-387-69033-9\_5},
}

@misc{malevergne_investigating_2002,
	title = {Investigating {Extreme} {Dependences}: {Concepts} and {Tools}},
	doi = {https://doi.org/10.2139/ssrn.303465},
	publisher = {SSRN},
	author = {Malevergne, Y. and Sornette, D.},
	year = {2002},
}

@incollection{baudin_openturns_2017,
	title = {{OpenTURNS}: {An} {Industrial} {Software} for {Uncertainty} {Quantification} in {Simulation}},
	isbn = {978-3-319-12384-4 978-3-319-12385-1},
	booktitle = {Handbook of {Uncertainty} {Quantification}},
	publisher = {Springer International Publishing},
	author = {Baudin, Michaël and Dutfoy, Anne and Iooss, Bertrand and Popelin, Anne-Laure},
	year = {2017},
	doi = {https://doi.org/10.1007/978-3-319-12385-1\_64},
	pages = {2001--2038},
}

@phdthesis{espoeys_optimisation_2025,
	title = {Optimisation multi-fidélité sous incertitudes, application à la conception de systèmes complexes},
	url = {https://theses.hal.science/tel-05129049/},
	school = {Université de Toulouse},
	author = {Espoeys, Romain},
	year = {2025},
}

@article{alves_note_2007,
	title = {A {Note} on {Second} {Order} {Conditions} in {Extreme} {Value} {Theory}: {Linking} {General} and {Heavy} {Tail} {Conditions}},
	issn = {2183-0371},
	doi = {https://doi.org/10.57805/revstat.v5i3.53},
	journal = {REVSTAT-Statistical Journal},
	author = {Alves, M. Isabel Fraga and Gomes, M. Ivette and de Haan, Laurens and Neves, Cláudia},
	year = {2007},
	pages = {285--304},
}

@inproceedings{trontin_description_2017,
	title = {Description and assessment of the new {ONERA} {2D} icing suite {IGLOO2D}},
	doi = {https://doi.org/10.2514/6.2017-3417},
	booktitle = {9th {AIAA} {Atmospheric} and {Space} {Environments} {Conference}},
	publisher = {American Institute of Aeronautics and Astronautics},
	author = {Trontin, Pierre and Blanchard, Ghislain and Kontogiannis, Alexandros and Villedieu, Philippe},
	year = {2017},
    pages={1--28},
}

@article{peherstorfer_survey_2018,
	title = {Survey of {Multifidelity} {Methods} in {Uncertainty} {Propagation}, {Inference}, and {Optimization}},
	volume = {60},
	issn = {0036-1445, 1095-7200},
	doi = {https://doi.org/10.1137/16m1082469},
	journal = {SIAM Review},
	author = {Peherstorfer, Benjamin and Willcox, Karen and Gunzburger, Max},
	year = {2018},
	pages = {550--591},
}

@article{ahmed_improved_2019,
	title = {Improved estimation of the extreme value index using related variables},
	volume = {22},
	issn = {1386-1999, 1572-915X},
	doi = {https://doi.org/10.1007/s10687-019-00358-y},
	journal = {Extremes},
	author = {Ahmed, Hanan and Einmahl, John H.J.},
	year = {2019},
	pages = {553--569},
}

@techreport{einmahl_variance-reduced_2024,
	title = {Variance-{Reduced} {Risk} {Inference} in {Semi}-{Supervised} {Settings}},
	author = {Einmahl, John H.J. and Peng, Liang},
	year = {2024},
    publisher = {{CentER}, {Center} for {Economic} {Research}},
    pages = {1--17},
    institution = {CentER, Center for Economic Research},
    address={}
}

@article{gorodetsky_generalized_2020,
	title = {A generalized approximate control variate framework for multifidelity uncertainty quantification},
	volume = {408},
	journal = {Journal of Computational Physics},
	author = {Gorodetsky, Alex A. and Geraci, Gianluca and Eldred, Michael S. and Jakeman, John D.},
	year = {2020},
	pages = {109257},
    doi={http://doi.org/10.1016/j.jcp.2020.109257}
}

@article{zhu_recent_2025,
	title = {Recent {Developments} on {Statistical} {Transfer} {Learning}},
	issn = {0306-7734, 1751-5823},
	doi = {http://doi.org/10.1111/insr.12613},
	journal = {International Statistical Review},
	author = {Zhu, Zhengyu and Yan, Yibo and Li, Gefei and Zhang, Riquan},
	year = {2025},
}

@article{bobbia_donsker_2025,
    author = {Bobbia, B. and Dombry, C. and Varron, D.},
    title = {A {Donsker} and {Glivenko}-{Cantelli} theorem for random measures linked to extreme value theory},
    journal = {{Scandinavian} {Journal} of {Statistics}},
    year = {2025},
    doi = {https://doi.org/10.1111/sjos.70007},
}

@article{bocquet_control_2025,
    title={Control variates for variance-reduced ratio of means estimators},
    author={Bocquet-Nouaille, Louison and Morio, J{\'e}r{\^o}me and Bobbia, Benjamin},
    journal={arXiv preprint},
    year={2025},
    doi={https://doi.org/10.48550/arXiv.2510.13504}
}

@incollection{owen_variance_2013,
	title = {Variance reduction},
	url = {https://artowen.su.domains/mc/},
	booktitle = {Monte {Carlo} theory, methods and examples},
	publisher = {{\textbackslash}url\{https://artowen.su.domains/mc/\}},
	author = {Owen, Art B.},
	year = {2013},
	pages = {28--33},
}

@article{caeiro_minimum_2020,
  title={Minimum-variance reduced-bias estimation of the extreme value index: A theoretical and empirical study},
  author={Caeiro, Frederico and Henriques-Rodrigues, L{\'\i}gia and Gomes, M Ivette and Cabral, Ivanilda},
  journal={Computational and Mathematical Methods},
  year={2020},
  publisher={Wiley Online Library},
  doi={https://doi.org/10.1002/cmm4.1101}
}

@article{gomes_new_2016,
  title={New reduced-bias estimators of a positive extreme value index},
  author={Gomes, M Ivette and Brilhante, M F{\'a}tima and Pestana, Dinis},
  journal={Communications in Statistics-Simulation and Computation},
  pages={833--862},
  year={2016},
  publisher={Taylor \& Francis},
  doi={https://doi.org/10.1080/03610918.2013.875567}
}

@article{cai_bias_2013,
  title={Bias correction in extreme value statistics with index around zero},
  author={Cai, Juan-Juan and {de Haan}, Laurens and Zhou, Chen},
  journal={Extremes},
  volume={16},
  number={2},
  pages={173--201},
  year={2013},
  publisher={Springer},
  doi={https://doi.org/10.1007/s10687-012-0158-x}
}

@Book{de_haan_regular_1975,
 title = {On regular variation and its application to the weak convergence of sample extremes},
 author = {{de Haan}, Laurens},
 publisher = {{Centrum} {Voor} {Wiskunde} en {Informatica}},
 year = 1975,
}

\end{document}